\documentclass[english,reprint,aps,prb,longbibliography]{revtex4-2}
\usepackage[T1]{fontenc}
\usepackage[latin9]{inputenc}
\usepackage{color}
\usepackage{babel}
\usepackage{amsmath}
\usepackage{amssymb}
\usepackage{graphicx}
\usepackage{setspace}
\usepackage{esint}
\usepackage[unicode=true,pdfusetitle,
 bookmarks=true,bookmarksnumbered=true,bookmarksopen=true,bookmarksopenlevel=1,
 breaklinks=false,pdfborder={0 0 0},pdfborderstyle={},backref=false,colorlinks=true]
 {hyperref}

\makeatletter

\newcommand{\lyxdot}{.}

\makeatother

\begin{document}
\global\long\def\ar{\mathbb{R}}%

\global\long\def\ci{\mathbb{C}}%

\global\long\def\a{\alpha}%

\global\long\def\b{\beta}%

\global\long\def\k{\kappa}%

\global\long\def\kev{\kappa_{\mathrm{ev}}}%

\global\long\def\imkev{\sqrt{\varphi}}%

\global\long\def\im#1{\mathrm{Im}\left(#1\right)}%

\global\long\def\re#1{\mathrm{Re}\left(#1\right)}%

\global\long\def\do{\partial}%

\title{Generalized topological bulk-edge correspondence in bulk-Hermitian
continuous systems with non-Hermitian boundary conditions}
\author{\singlespacing{}Orr Rapoport}
\email{orrrapoport@mail.tau.ac.il}

\author{Moshe Goldstein}
\affiliation{\textit{\small{}Raymond and Beverly Sackler School of Physics and
Astronomy, Tel Aviv University, Tel Aviv, 6997801, Israel}}
\begin{abstract}
The bulk-edge correspondence (BEC) is the hallmark of topological
systems. In continuous (non-lattice) Hermitian systems with an unbounded
wave vector, it was recently shown that the BEC of Chern insulators
is modified. How would it be further affected in non-Hermitian systems,
experiencing loss and/or gain? In this work, we take the first step
in this direction, by studying a bulk-Hermitian continuous system
with non-Hermitian boundary conditions. We find in this case that
edge modes emerge at the roots of the scattering matrix, as opposed
to the Hermitian case, where they emerge at its poles (or, more accurately,
coalescence of roots and poles). This entails a nontrivial modification
to the relative Levinson's theorem. We then show that the topological
structure remains the same as in the Hermitian case, and the generalized
BEC holds, provided one employs appropriately modified contours in
the wave-vector plane so that the scattering matrix phase winding
counts the edge modes correctly. We exemplify all this using a paradigmatic
model of waves in a shallow ocean or active systems in the presence
of odd viscosity, as well as 2D electron gas with Hall viscosity.
We use this opportunity to examine the case of large odd viscosity,
where the scattering matrix becomes $2\times2$, which has not been
discussed in previous works on the Hermitian generalized BEC.
\end{abstract}
\maketitle

\section{Introduction}

The study of topological band structure has been of great theoretical
and experimental interest in recent years, with applications in various
fields in physics, including solid state physics \citep{Hasan2010,Bansil2016,Qi2011,Haldane1988},
photonics \citep{Khanikaev2013,Lu2014,Ozawa2019a,Harari2018,Bandres2018},
mechanical systems \citep{Xin2020,Ma2019,Zhang2018,Kane2013,Prodan2009,He2016},
electronic circuits \citep{Lee}, and active matter \citep{Shankar2022,Shankar2017,Souslov2017,Souslov2019}.
At the heart of this field lies the bulk-edge correspondence (BEC),
which is the relation between a topological invariant calculated on
the bulk, and the number of available topologically-protected edge
modes \citep{Thouless1982,Hatsugai1993a,Hatsugai1993b}. Even though
topological effects are usually considered on lattice models, there
are unique consequences for continuous (nonlattice) systems \citep{Bal2019,Tauber2018,Souslov2019,Tauber2020,Silveirinha2019,VanMechelen2018,VanMechelen2021}.
In particular, it has been shown in a system with nontrivial Chern
number that the effect of a noncompact space for the wave vector $\mathbf{k}$
leads to a generalized BEC: One must account for the behavior at $\left|\mathbf{k}\right|\to\infty$
in order for the BEC to be obeyed \citep{Graf2020}. 

Another central concept in recent studies is non-Hermitian systems,
appropriate for classical wave systems with loss and/or gain, as well
as an effective description of open quantum systems \citep{Rotter2009,Yokomizo2019,Ashida2020,Bender2007}.
In the context of topology, recent studies have shown that the BEC
is affected by non-Hermiticity \citep{Bergholtz2021,Shen2018a,Gong2018,Esaki2011,Kunst2018}
in several ways, the most well-known of which is the non-Hermitian
skin effect, which violates the BEC \citep{Lee2016,MartinezAlvarez2018,Yao2018a,Yokomizo2020,Xiong2018,Leykam2017,Kunst2019,Lee2019,Imura2019,Longhi2021}.

It is thus natural to ask how topology is affected by non-Hermiticity
in continuous non-lattice systems. We take here the first step in
bringing together these two concepts, by considering the simplest
such system, where the bulk remains Hermitian, and non-Hermiticity
is obtained only from the choice of a certain boundary condition.
For this end we study the shallow-water model \citep{Vallis2017},
which originally describes shallow ocean waves, but also applies to
active matter systems in the presence of odd viscosity \citep{Souslov2019},
as well as 2D electron gas in the presence of a magnetic field and
with Hall viscosity \citep{Cohen2018b}. 

In this work, we build upon the Hermitian generalized BEC of Chern
insulators introduced in Ref. \citep{Graf2020}. There, using a relative
version of Levinson's theorem \citep{Graf2012}, the BEC is amended
to account for the behavior at $\left|\mathbf{k}\right|\to\infty$.
The main idea in using Levinson's theorem in this context, is that
the edge states can be thought of as bound states in terms of the
motion perpendicular to the edge, while the parallel wavevector acts
as a parameter. The theorem states that since poles (or, more accurately,
singularities where roots and poles coincide) of the scattering matrix
of bulk modes at the edge correspond to the emergence of bound states,
then recording the phase of the scattering matrix along a contour
near these poles must count the number of edge states, as the phase
changes by $2\pi$ near each of them --- increasing for modes emerging
from the bottom of the upper bulk band and decreasing for modes merging
into it. Now, the bulk Chern number can be related to the phase accumulation
of the scattering matrix along a closed contour in the $\mathbf{k}$
plane. The latter is composed of two contributions: a finite-$\mathbf{k}$
part counting all edge modes by the relative Levinson's theorem, as
just explained, and an additional contribution from the (unbounded)
band top, namely, at $\left|\mathbf{k}\right|\to\infty$. This last
contribution amends the apparent mismatch in the BEC \footnote{A different approach has recently been presented in Ref. \citep{Tauber2022b}.
There the spectral flow of the edge modes as a function of an appropriate
parametrization of the family of boundary conditions (\ref{eq:boundary_cond})
was considered, leading to a BEC which in a sense averages over this
family. However, in many situations a fixed boundary condition is
considered, with no straightforward generalization to an appropriate
family, in particular non-Hermitian ones. Hence we prefer to study
fixed boundary conditions in the non-Hermitian case}.

Our main result is demonstrating and explaining a generalized BEC
for the non-Hermitian edge problem we consider, using a modified version
of the relative Levinson's theorem. In the non-Hermitian case, since
edge mode dispersions have nonzero imaginary parts, they intersect
(merge into/emerge from) the bulk bands away from their top/bottom.
This is related to the separation of the roots and poles of the scattering
matrix, so that \emph{edge mode intersections with the bulk bands
now occur at roots of the scattering matrix}. Thus, the claim in the
non-Hermitian relative Levinson's theorem is that the phase of the
scattering matrix records edge mode intersections with the bulk bands,
provided one chooses a contour going above these zeros. 

\paragraph*{Non-Hermitian relative Levinson's theorem:\label{par:Modified-relative-Levinson's-thm}}

Let $\left(K_{x,\alpha},K_{y,\alpha}\right)\in\ar\times\ar_{+},\alpha=1,2,...,N$,
be the momenta at which edge mode dispersions intersect with the bulk
band (merge into it or emerge from it), and let $k_{y_{0}}>\max_{\a}\left(K_{y,\alpha}\right)$.
Then
\begin{equation}
\lim_{\substack{k_{x_{1}}\rightarrow-\infty\\
k_{x_{2}}\rightarrow\infty
}
}\arg\left[S\left(k_{x},k_{y_{0}}\right)\right]|_{k_{x_{1}}}^{k_{x_{2}}}=2\pi n,\label{eq:modified_levinson_thm_orr}
\end{equation}
where $S$ is the scattering matrix of the bulk eigenmodes at the
boundary, and $n$ is the signed number of edge modes emerging from
($+$) and merging into $(-$) the upper bulk band at its bottom.

Adding the band-top contribution as in the Hermitian case, we will
arrive at the non-Hermitian generalized BEC:

\paragraph*{Non-Hermitian generalized bulk-edge correspondence:\label{par:Generalized-bulk-edge-correspond}}

\begin{equation}
C_{+}=\mathrm{Ind}\left(S\left(\left|\mathbf{k}\right|\rightarrow\infty\right)\right)+n,\label{eq:NH_generalized_BEC}
\end{equation}
where $C_{+}$ is the first Chern number of the upper bulk band, $\mathrm{Ind}\left(S\right)$
is the winding number of $S$ over a contour $\gamma$, $\mathrm{Ind}\left(S\right)=\frac{1}{2\pi i}\int_{\gamma}S^{-1}dS$,
and by $\mathrm{Ind}\left(S\left(\left|\mathbf{k}\right|\rightarrow\infty\right)\right)$
we mean $\gamma$ is taken as a contour along the top of the (unbounded)
upper bulk band.

The above statement of the generalized BEC is exactly as in theorem
2.9 in Ref. \citep{Graf2020}. The difference lies in the modification
to the relative Levinson's theorem, which is a direct generalization
of theorem 3.2 in Ref. \citep{Graf2020}. The main difference that
should be noted is that one must take a contour ``above'' the roots,
i.e., $k_{y_{0}}>K_{y,\alpha}$, due to the separation of roots and
poles of the scattering matrix. This causes the phase accumulation
to be smeared, and thus equality is achieved only in the limit of
covering the entire real $k_{x}$ line, unlike the Hermitian case
where it is enough to take finite $k_{x_{1}},k_{x_{2}}$ large enough
so that all $K_{x,\a}$ are in the domain $\left(k_{x_{1}},k_{x_{2}}\right)$,
as well as the limit $k_{y}\rightarrow0$ (approaching the bottom
of the upper bulk band).

In this work, we examine in addition the case of large odd viscosity
for both the Hermitian and non-Hermitian edge systems, which has not
been discussed in previous works on the Hermitian case. We show that
large odd viscosity leads to a $2\times2$ scattering matrix instead
of a scalar, necessitating generalized definitions and methods of
calculation, but ultimately resulting in the same generalized BEC.
In particular, the phase used in the non-Hermitian relative Levinson's
theorem is that of $\det S$. 

The rest of this paper is organized as follows. In Sec. \ref{sec:Dissipationless-model},
we set the stage by introducing the shallow-water model \citep{Vallis2017},
which also describes active systems \citep{Souslov2019} or 2D electron
gas in the presence of a magnetic field \citep{Cohen2018b}, and which
we will use throughout to exemplify our claims. We will also review
the general method of using the scattering matrix to account for the
BEC mismatch in the Hermitian case, mainly repeating known results.
In Sec. \ref{sec:Non-hermitian-edge-problem}, we introduce a family
of non-Hermitian boundary conditions, and show their effect on the
scattering matrix, especially that now its roots are related to edge
mode emergence. As a result, the relative Levinson's theorem is modified.
We prove the theorem using a topological argument and use it to characterize
the resulting non-Hermitian generalized BEC. We verify all these results
numerically. In Sec. \ref{sec:large-odd-viscosity}, we consider the
case of large odd viscosity for both Hermitian and non-Hermitian systems,
which has not been treated before in works regarding the Hermitian
case. We finish in Sec. \ref{sec:Discussion-and-conclusions} with
discussion and conclusions. We defer some calculations and remarks
to the appendices. In Appendix \ref{sec:App_neg_oddvisc} we briefly
present the case of negative odd viscosity (namely, with opposite
sign with respect to the Coriolis term), where, for both Hermitian
and non-Hermitian cases, although bulk topology is trivial, edge modes
may appear, and are accounted for by the behavior of $S$ at $\left|\mathbf{k}\right|\to\infty$,
in accordance with the generalized BEC. In Appendix \ref{sec:App.proof-NHBCisdissipative},
we prove that the family of Hermitian boundary conditions we consider
indeed preserves Hermiticity for the edge problem, whereas the non-Hermitian
boundary conditions break it, and specifically cause dissipation.
In Appendix \ref{sec:App-detS=00003D1_interpolated_bc}, we show that
in the non-Hermitian system with $1\times1$ scattering matrix, $\left|S\right|<1$
in the domain of the wavevector plane where it is well-defined, as
appropriate for a dissipative boundary condition. In Appendix \ref{sec:APP-Analytic-continuation},
we discuss the ``anomaly of Levinson's theorem'' at $\left|\mathbf{k}\right|\to\infty$,
observed in Ref. \citep{Graf2020}, and show analytically that it
occurs for the strictly non-Hermitian system, but not for the no-slip
condition. In Appendix \ref{sec:App-Equivalence-of-numerical-methods},
we discuss two possible numerical evaluation methods for the phase
of the scattering matrix and prove their equivalence. In Appendix
\ref{sec:app_detS=00003D1_bigfnu_interpolated}, we extend the results
of Appendix \ref{sec:App-detS=00003D1_interpolated_bc} to the non-Hermitian
system with large odd viscosity, where $S$ is a $2\times2$ matrix,
and show that $\left|\det S\right|<1$ in the domain of the wavevector
plane where it is well-defined, again in accordance with loss rather
than gain. In Appendix \ref{sec:app-Topographic-map-of_dets}, we
give more details on the separation of roots and poles of the scattering
matrix in the non-Hermitian system with large odd viscosity.

\begin{figure*}[t]
\includegraphics[scale=0.5]{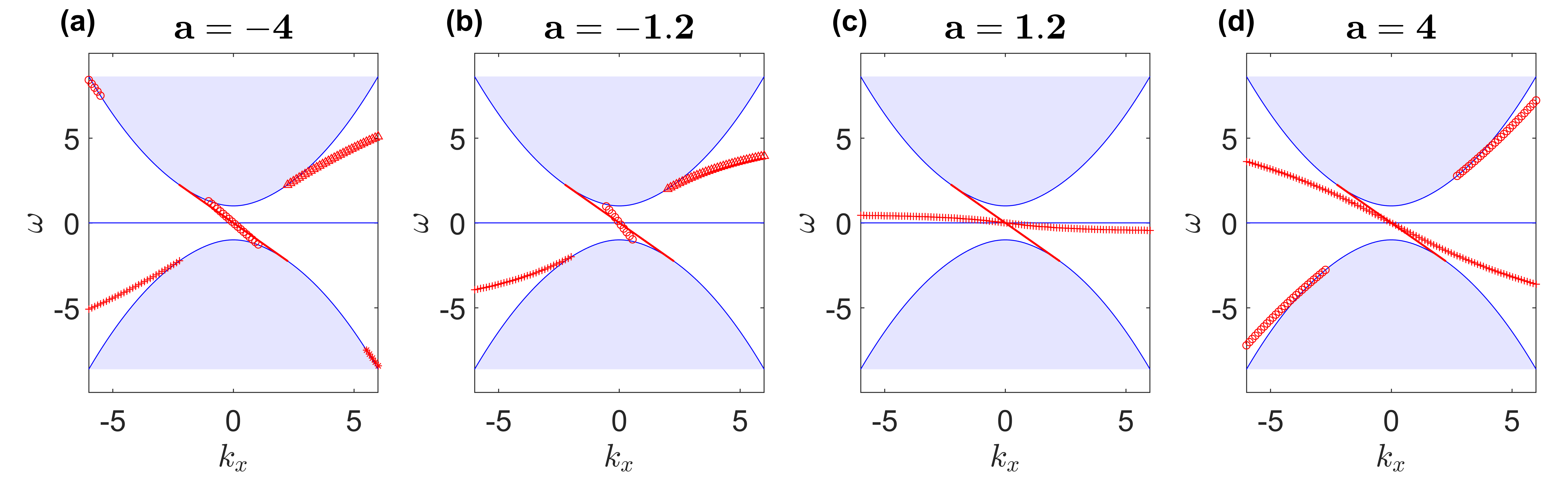}

\caption{\label{fig:Dispersion-multi-adep}Dispersion plots for different values
of $a$, for the parameter-dependent Hermitian boundary condition
(\ref{eq:boundary_cond}). Filled in light blue are the bulk bands,
projected on the $\left(k_{x},\omega\right)$ plane. The borders of
the bulk bands, obtained for $k_{y}=0$ {[}see Eq. (\ref{eq:original_dispersion}){]},
are in blue. Here $f=1,\nu=0.2$, similarly to Fig. 1 in Ref. \citep{Graf2020}.
We count the number of edge modes by intersections with the upper
bulk band (positive for a mode emerging from the bulk band and negative
for a mode merging with it). For $\left|a\right|>\sqrt{2}$, one finds
two edge modes, fitting the usual BEC, but for $-\sqrt{2}<a<0$, three
edge modes are seen, and for $0<a<\sqrt{2}$ only one. All have the
parameter-independent mode $\omega=-k_{x}$, and all edge mode dispersion
intersections with the upper band occur at its bottom, namely, with
$k_{y}=0$.}
\end{figure*}

\section{Recap of the Hermitian shallow-water model\label{sec:Dissipationless-model}}

In this section we present the fully Hermitian version of the model
and the generalized BEC, following Ref. \citep{Graf2020}. This recap
acts as a preparation, and basis for comparison with our results for
non-Hermitian systems, presented in later sections.

Consider the shallow-water model: Assume the typical wavelength in
the fluid is much larger than its depth, so pressure is approximately
hydrostatic and the velocity can be averaged over columns of fluid.
Thus the problem is effectively two-dimensional. We denote by $h$
the total height of the surface, by $H$ the mean height of the surface
with respect to the bottom, by $\eta$ the height of the surface relative
to its mean, i.e., $\eta=h/H$, and by $\mathbf{V}=\left(U,V\right)$
the two-component velocity field. The three quantities $\eta,U,V$
are all functions of the two-dimensional coordinates $\mathbf{X}=\left(X,Y\right)\in\mathbb{R}^{2}$
and time $t\in\mathbb{R}$. Assume an incompressible and homogeneous
fluid. Under these assumptions, the linearized shallow-water equations
for a rotating frame are found from mass and momentum conservation
\citep{Vallis2017},
\begin{align}
\do_{t}h & =-H\left(\do_{X}U+\do_{Y}V\right),\nonumber \\
\do_{t}U & =-g\do_{X}h-\left(f+\tilde{\nu}\nabla_{\mathbf{X}}^{2}\right)V,\nonumber \\
\do_{t}V & =-g\do_{Y}h+\left(f+\tilde{\nu}\nabla_{\mathbf{X}}^{2}\right)U,
\end{align}
where $f$ is the Coriolis term, $g$ is the gravitational acceleration,
and $\tilde{\nu}$ is the kinematic odd viscosity term \citep{Avron1998},
i.e, per mass density. Note that we neglect dissipative terms, and
specifically even viscosity. We divide all equations by $\sqrt{gH}$
and define dimensionless velocities $\mathbf{v}=\mathbf{V}/\sqrt{gH}$
to find
\begin{align}
\do_{t}\eta & =-\sqrt{gH}\left(\do_{X}u+\do_{Y}v\right),\nonumber \\
\do_{t}u & =-\sqrt{gH}\do_{X}\eta-\left(f+\tilde{\nu}\nabla_{\mathbf{X}}^{2}\right)v,\nonumber \\
\do_{t}v & =-\sqrt{gH}\do_{Y}\eta+\left(f+\tilde{\nu}\nabla_{\mathbf{X}}^{2}\right)u.
\end{align}
We thus see that we can redefine the coordinates $\mathbf{x}=\mathbf{X}/\sqrt{gH}$
so that time and space are in the same units. We finally redefine
$\nu=\tilde{\nu}/gH$ and write the system of PDEs for the problem
with dimensionless $\eta,u,v$, which we will use throughout,
\begin{align}
\partial_{t}\eta & =-\partial_{x}u-\partial_{y}v,\nonumber \\
\partial_{t}u & =-\partial_{x}\eta-\left(f+\nu\nabla^{2}\right)v,\nonumber \\
\partial_{t}v & =-\partial_{y}\eta+\left(f+\nu\nabla^{2}\right)u.\label{eq:orig_PDEs}
\end{align}
We assume throughout positive $f,\nu$, and for this and the next
section we also assume $f\nu<1/4$, meaning not-too-large odd viscosity;
the case $f\nu>1/4$ has not been addressed before even in the fully
Hermitian case, and we will explore it later on, in Sec. \ref{sec:large-odd-viscosity}.
The case of negative $f,\nu$ has the same topological structure as
will be presented below, up to inversion of sign ($C_{+}=-2$). The
case of $f\nu<0$ , namely, one of them is negative and the other
positive, leads to trivial bulk topology ($C_{+}=0$), and yet edge
modes may appear, following the generalized BEC. This is discussed
further in Appendix \ref{sec:App_neg_oddvisc}.

A few more words on the model are due before moving forward. This
model was used to demonstrate topologically-protected modes in the
context of geophysical waves in Ref. \citep{Tauber2020}, where small
odd viscosity is required for regularization, yet is less physically
motivated. This same model also applies for active chiral materials
\citep{Souslov2019}, and for 2D electron gas in a magnetic field
\citep{Cohen2018b}, where for both the odd viscosity term is indeed
physical, and can attain large values. In the electronic system, odd
viscosity is manifested by the Hall viscosity, and Coriolis is replaced
by a perpendicular magnetic field, which breaks time reversal symmetry
\citep{Avron1995,Bradlyn2012}.

\subsection{Bulk eigenmodes}

The system (\ref{eq:orig_PDEs}) is analogous to the time-dependent
Schrödinger equation
\begin{align}
i\partial_{t}\psi & =H\psi,\,\psi=\left(\begin{array}{c}
\eta\\
u\\
v
\end{array}\right),\nonumber \\
H & =\begin{pmatrix}0 & p_{x} & p_{y}\\
p_{x} & 0 & -i\left(f-\nu\mathbf{p}^{2}\right)\\
p_{y} & i\left(f-\nu\mathbf{p}^{2}\right) & 0
\end{pmatrix},\label{eq:orig_hamiltonian}
\end{align}
with $p_{x}=-i\partial_{x},p_{y}=-i\partial_{y},p^{2}=p_{x}^{2}+p_{y}^{2}=-\nabla^{2}$.
In the bulk domain, $L^{2}(\mathbb{R}^{2})^{\otimes3}$, this Hamiltonian
is self-adjoint. Since it is also translation-invariant in both directions,
we have the bulk eigenmodes
\begin{equation}
\psi=\hat{\psi}e^{i\left(k_{x}x+k_{y}y-\omega t\right)},\,\hat{\psi}=\begin{pmatrix}\hat{\eta}\\
\hat{u}\\
\hat{v}
\end{pmatrix},
\end{equation}
with $\left(k_{x},k_{y}\right)\in\mathbb{R}^{2},\mathbf{k}^{2}=k_{x}^{2}+k_{y}^{2}$,
leading to the eigenvalue problem
\begin{equation}
\hat{H}\hat{\psi}=\omega\hat{\psi},\,\hat{H}=\begin{pmatrix}0 & k_{x} & k_{y}\\
k_{x} & 0 & -i\left(f-\nu\mathbf{k}^{2}\right)\\
k_{y} & i\left(f-\nu\mathbf{k}^{2}\right) & 0
\end{pmatrix}.\label{eq:orig_hamiltonian_eigenvalue_prob}
\end{equation}
This system admits three eigenvalues for each $\left(k_{x},k_{y}\right)$,
meaning three bands in the bulk dispersion
\begin{equation}
\omega_{\pm}\left(\mathbf{k}\right)=\pm\sqrt{\mathbf{k}^{2}+\left(f-\nu\mathbf{k}^{2}\right)^{2}},\ \omega_{0}\left(\mathbf{k}\right)=0.\label{eq:original_dispersion}
\end{equation}
Since $\omega_{+}\left(k_{x},k_{y}\right)>\omega_{+}\left(k_{x},0\right)$
for all $\left(k_{x},k_{y}\right)$, the projection of the bulk dispersion
on $\left(k_{x},\omega\right)$ is given by $\omega_{+}\left(k_{x},0\right)$
and the area above it (see Fig. \ref{fig:Dispersion-multi-adep}).

For later use, the eigenvectors corresponding to the upper band are
given by
\begin{align}
\hat{\psi}^{\infty}\left(k_{x},k_{y}\right) & =\frac{1}{\sqrt{2}}\frac{1}{k_{x}-ik_{y}}\begin{pmatrix}\mathbf{k}^{2}/\omega^{2}\\
k_{x}-ik_{y}q\\
k_{y}+ik_{x}q
\end{pmatrix},\nonumber \\
q\left(\mathbf{k}\right) & =\frac{f-\nu\mathbf{k}^{2}}{\omega},\ \omega=\omega_{+}\left(\mathbf{k}\right).\label{eq:psi_inf_sol_q}
\end{align}
Note that $q\rightarrow1$ for $\left|\mathbf{k}\right|\rightarrow0$,
and $q\rightarrow-1$ for $\left|\mathbf{k}\right|\to\infty$, making
the solution at $\infty$ dependent on the angle of the wavevector
$\mathbf{k}$, and thus regular anywhere but $\infty$. In order to
investigate the solution around $\infty$ one can move the singularity
to an arbitrary finite complex point $\zeta$:
\begin{equation}
\hat{\psi}^{\zeta}=t_{\infty}^{\zeta}\hat{\psi}^{\infty},\ t_{\infty}^{\zeta}\left(z\right)=\frac{\bar{z}-\bar{\zeta}}{z-\zeta},\label{eq:tzeta}
\end{equation}
where $z=k_{x}+ik_{y}$ in this context. For abstract proofs $\zeta=i\zeta_{y}$
with real positive $\zeta_{y}$ is useful, and for numerical calculations
around $\left|\mathbf{k}\right|\to\infty$, $\zeta=0$ suffices.

From the bulk eigenmodes (\ref{eq:psi_inf_sol_q}) we see there is
no nonvanishing, regular global eigensection, namely, at least two
sections are required for the bulk eigenmodes to be well-defined on
the compactified wavevector plane $\ar^{2}\cup\left\{ \infty\right\} \cong S^{2}$.
Thus the bulk dispersion bands carry nontrivial topology. It has been
shown {[}see Eq. (\ref{eq:Chern_in_SWM}) below{]} that the bulk Chern
number for the upper band, denoted $C_{+}$, equals $2$, assuming
positive $f,\nu$. Since $C_{+}$ is a bulk topological invariant,
\emph{it is} \emph{independent of the boundary conditions to be discussed
next}.

\begin{figure}[t]
\includegraphics[scale=0.34]{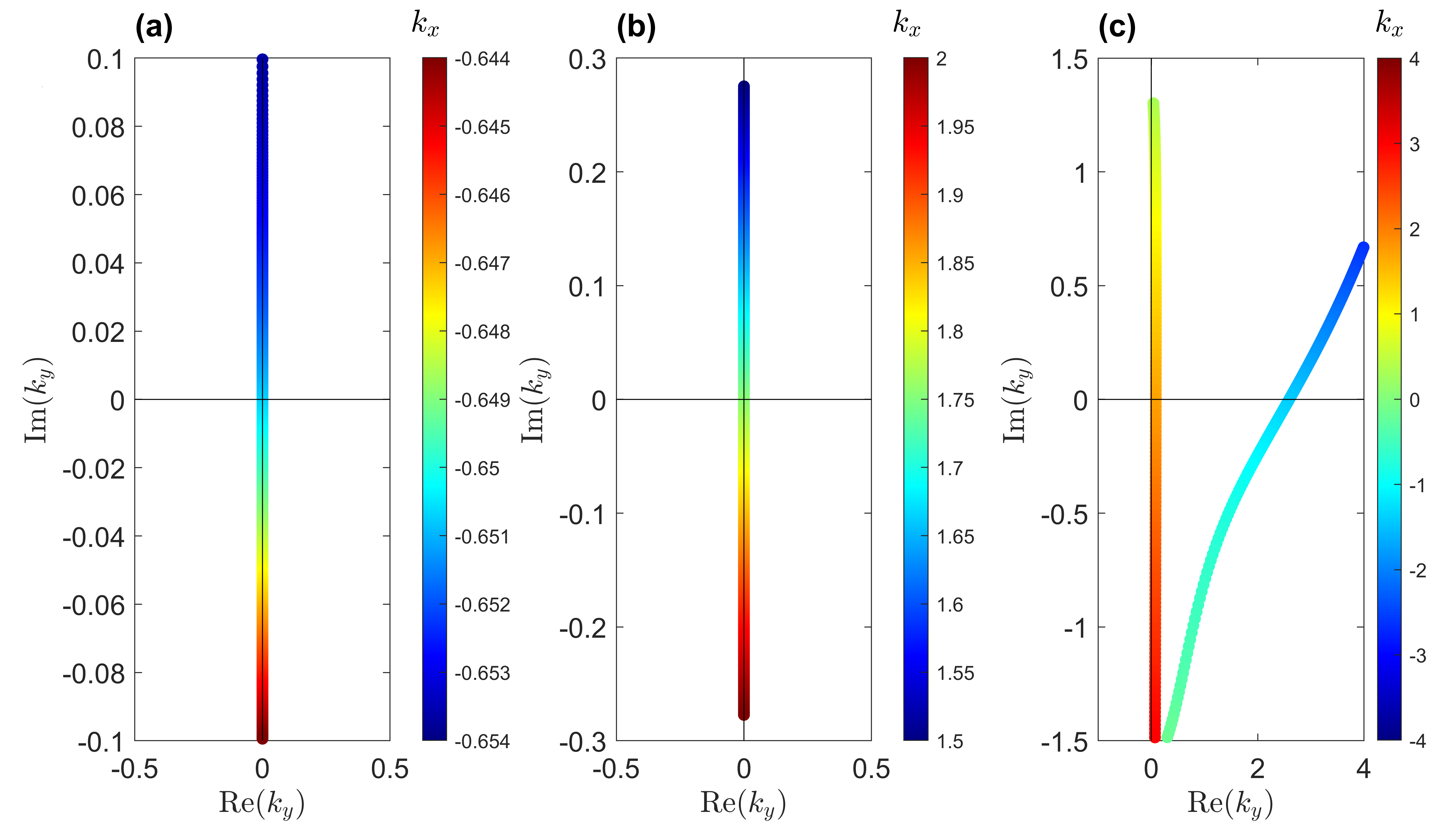}

\caption{\label{fig:kycomplex_Sroots}Roots of $S_{\infty}$ in the complex
$k_{y}$ plane, with $k_{x}$ as a parameter (shown by color scale),
calculated numerically for $f=1,\nu=0.2$. We depict only the solutions
crossing $\protect\im{k_{y}}=0$, and exclude the parameter-independent
solution $\omega=-k_{x}$. {[}(a) and (b){]} The Hermitian case with
the parameter-dependent boundary condition (\ref{eq:boundary_cond}),
with $a=-1.2$. As discussed in Sec. \ref{subsec:Generalized-BEC}
these \textquotedblleft roots\textquotedblright{} are actually singularities
in this case. The two real solutions $K_{x,1}\approx-0.65,K_{y,1}=0$
in (a) and $K_{x,2}\approx1.75,K_{y,2}=0$ in (b), are separated for
clarity. (c) The non-Hermitian case with the partial-slip boundary
condition (\ref{eq:explicit_interpolated_BC}), with $\tilde{a}=10$.
Two real roots are found with $\protect\re{K_{y}}>0$: $\left(K_{x,1},K_{y,1}\right)\approx\left(1.7,0.09\right)$,$\left(K_{x,2},K_{y,2}\right)\approx\left(-1.43,2.6\right)$.
All the roots shown move as $k_{x}$ grows from $\protect\im{K_{y}}>0$
to $\protect\im{K_{y}}<0$, in agreement with the discussion in Sec.
\ref{subsec:Generalized-BEC}. In all cases for a root at $K_{y}$
there is a corresponding pole at $-K_{y}$, see Eqs. (\ref{eq:gzeta_and_Szeta_def}),(\ref{eq:interpolated_Smatrix}).}
\end{figure}

\subsection{Edge eigenmodes\label{subsec:Hermitian_edge-eigenmodes}}

We use the same Hamiltonian (\ref{eq:orig_hamiltonian}), but now
the problem is defined in the upper half-plane $\left(x,y\right)\in\ar\times\ar_{+}$,
i.e., there is a boundary along the $x$ axis. Following Ref. \citep{Graf2020},
we define the parameter-dependent boundary condition:
\begin{equation}
v|_{y=0}=0,\left(\partial_{x}u+a\partial_{y}v\right)|_{y=0}=0,\label{eq:boundary_cond}
\end{equation}
where $a\in\mathbb{R}$ is a parameter. The first part of this boundary
condition means that velocity perpendicular to the surface vanishes.
The second part is a family of conditions that will allow to explore
the manifestation of the generalized BEC. For $a=1$ it means that
$\eta$ is fixed at the boundary {[}due to the first of Eqs. (\ref{eq:orig_PDEs}){]},
and for $a=-1$ it reduces to the no-stress condition $T_{xy}|_{y=0}=0$,
meaning no parallel force acts on the fluid at the boundary. As shown
in Appendix \ref{sec:App.proof-NHBCisdissipative}, the system is
Hermitian with these boundary conditions for any value of $a$.

We now present the method to find the edge mode dispersion. Since
the half-plane maintains translation invariance in the $x$ direction,
we have solutions of the form
\begin{equation}
\psi=\tilde{\psi}e^{i(k_{x}x-\omega t)},\tilde{\psi}=\begin{pmatrix}\tilde{\eta}\\
\tilde{u}\\
\tilde{v}
\end{pmatrix}.\label{eq:bound_psi_form}
\end{equation}
Hence the edge dispersion will be discrete eigenvalue solutions to
$H\left(k_{x}\right)$. We use an ansatz of states bound to the edge,
\begin{equation}
\tilde{\eta}=\eta_{0}e^{-\kappa y},\tilde{u}=u_{0}e^{-\kappa y},\tilde{v}=v_{0}e^{-\kappa y},\label{eq:uv_bound_form}
\end{equation}
meaning we require $\re{\k}>0$. This is a more physically-motivated
approach than directly solving the general eigenvalue equation and
finding retroactively which solution are edge modes (as in Ref. \citep{Tauber2020}),
but of course leads to the same results. Plugging this form into Eqs.
(\ref{eq:orig_PDEs}), we can substitute $\eta_{0}$ and find two
equations for $u_{0},v_{0}$:
\begin{align}
\frac{i}{\omega}\left(\omega^{2}-k_{x}^{2}\right)u_{0} & =\left(f+\nu\left(\kappa^{2}-k_{x}^{2}\right)-\frac{k_{x}\kappa}{\omega}\right)v_{0},\nonumber \\
-\frac{i}{\omega}\left(\omega^{2}+\kappa^{2}\right)v_{0} & =\left(f+\nu\left(\kappa^{2}-k_{x}^{2}\right)+\frac{k_{x}\kappa}{\omega}\right)u_{0}.\label{eq:uv_eqs}
\end{align}
Assume $u_{0},v_{0}\ne0$ (the case where one of them vanishes will
be discussed later on). A nontrivial solution to the above system
of linear equations in $u_{0},v_{0}$ requires equating the determinant
of the coefficients to zero, which results in the equation:
\begin{multline}
\frac{1}{\omega^{2}}\left(\omega^{2}-k_{x}^{2}\right)\left(\omega^{2}+\kappa^{2}\right)=\\
\left(f+\nu\left(\kappa^{2}-k_{x}^{2}\right)+\frac{k_{x}\kappa}{\omega}\right)\left(f+\nu\left(\kappa^{2}-k_{x}^{2}\right)-\frac{k_{x}\kappa}{\omega}\right).\label{eq:uv_bound_zero_det}
\end{multline}
This is a polynomial of degree $4$ in $\kappa$. Under the assumption
$f\nu<1/4$ we have $\kappa\in\mathbb{R}$ for all four solutions
and for all $k_{x}$. Since we are interested in bound edge modes,
we only take solutions with a positive overall sign,
\begin{equation}
\kappa_{1,2}=+\sqrt{k_{x}^{2}+\frac{1-2f\nu\pm\sqrt{1+4\nu\left(\nu\omega^{2}-f\right)}}{2\nu^{2}}}.\label{eq:kappa_def}
\end{equation}
We see that exactly two values of $\k$ are positive, and two are
negative, independently of the boundary condition. We can now plug
$\kappa_{j}$ back into Eqs. (\ref{eq:uv_eqs}) in order to obtain
a relation between $u_{0},v_{0}$,
\begin{equation}
u_{0}=\frac{\omega\left(f+\nu\left(\kappa_{j}^{2}-k_{x}^{2}\right)\right)-k_{x}\kappa_{j}}{i\left(\omega^{2}-k_{x}^{2}\right)}\text{\ensuremath{\equiv}}\lambda_{j}v_{0}.\label{eq:lambda_def}
\end{equation}
In order to use the boundary condition, we construct a superposition
of the solutions with the two values of $\kappa$,
\begin{equation}
\tilde{\psi}=\begin{pmatrix}\tilde{u}\\
\tilde{v}
\end{pmatrix}=A\begin{pmatrix}\lambda_{1}\\
1
\end{pmatrix}e^{-\kappa_{1}y}+B\begin{pmatrix}\lambda_{2}\\
1
\end{pmatrix}e^{-\kappa_{2}y}.\label{eq:edge_modes_sp}
\end{equation}
Plugging this solution into Eqs. (\ref{eq:boundary_cond}) gives a
linear system of equations, this time in the coefficients $A,B$,
which has a unique nontrivial solution if its matrix of coefficients
is non-invertible. Thus, equating the determinant to zero yields the
third desired equation,
\begin{equation}
a\left(\omega^{2}-k_{x}^{2}\right)=k_{x}\omega\nu\left(\kappa_{1}+\kappa_{2}\right)-k_{x}^{2}.\label{eq:omega_eq_with_kappa}
\end{equation}
We now have three equations, (\ref{eq:kappa_def}),(\ref{eq:omega_eq_with_kappa}),
tying together $\kappa_{1},\kappa_{2},\omega$, where of course $\kappa_{1}$
and $\kappa_{2}$ appear symmetrically. It should be noted that the
boundary conditions (\ref{eq:boundary_cond}) cannot be met for a
single edge mode, i.e., a superposition of at least two modes is required.
On the other hand, as seen from in Eq. (\ref{eq:kappa_def}), only
two values of $\k$ have the correct sign, independently of the boundary
condition. Taking the square of Eq. (\ref{eq:omega_eq_with_kappa})
and plugging in both $\kappa_{1},\kappa_{2}$ results in a polynomial
equation of degree 4 in $\omega^{2}$
\begin{widetext}
\begin{align}
0 & =a^{4}\omega^{8}+2k_{x}^{2}\left[-2a^{3}(a-1)+2\nu^{3}k_{x}^{2}-a^{2}(1-2f\nu+2\nu^{2}k_{x}^{2})\right]\omega^{6}\nonumber \\
 & +k_{x}^{4}\left[6a^{2}(a-1)^{2}+1-4f\nu+4a(a-1)(1-2f\nu+2\nu^{2}k_{x}^{2})\right]\omega^{4}\nonumber \\
 & +2k_{x}^{6}\left[-2a\left(a-1\right)^{3}-\left(a-1\right)^{2}(1-2f\nu+2\nu^{2}k_{x}^{2})\right]\omega^{2}+k_{x}^{8}(a-1)^{4}.
\end{align}
\end{widetext}

\noindent Of course by taking the square we have added artificial
solutions, so after finding the roots numerically we need to check
their validity by plugging them into Eq. (\ref{eq:omega_eq_with_kappa})
and verifying that $\re{\kappa_{j}}>0$.

We shortly address the cases either of $u_{0}=0$ or $v_{0}=0$. We
find that only $v_{0}=0$ complies with the boundary condition, and
leads to $\omega=-k_{x}$, which is the well-known Kelvin wave. The
requirement of $\re{\k_{j}}>0$ leads to the constraint that this
mode is defined only for $k_{x}\in\left[-k_{0},k_{0}\right],k_{0}\equiv\sqrt{f/\nu}$
(see Ref. \citep{Tauber2020}). We will see this parameter-independent
mode persists for all following cases, independent of the parameters
of the boundary conditions. Here, for (\ref{eq:boundary_cond}), we
see that $v_{0}=0$ everywhere means that the boundary condition reduces
to $\do_{x}u|_{y=0}=0$, and indeed any mode that obeys this condition
is independent of the parameter $a$.

\begin{figure*}[t]
\includegraphics[scale=0.42]{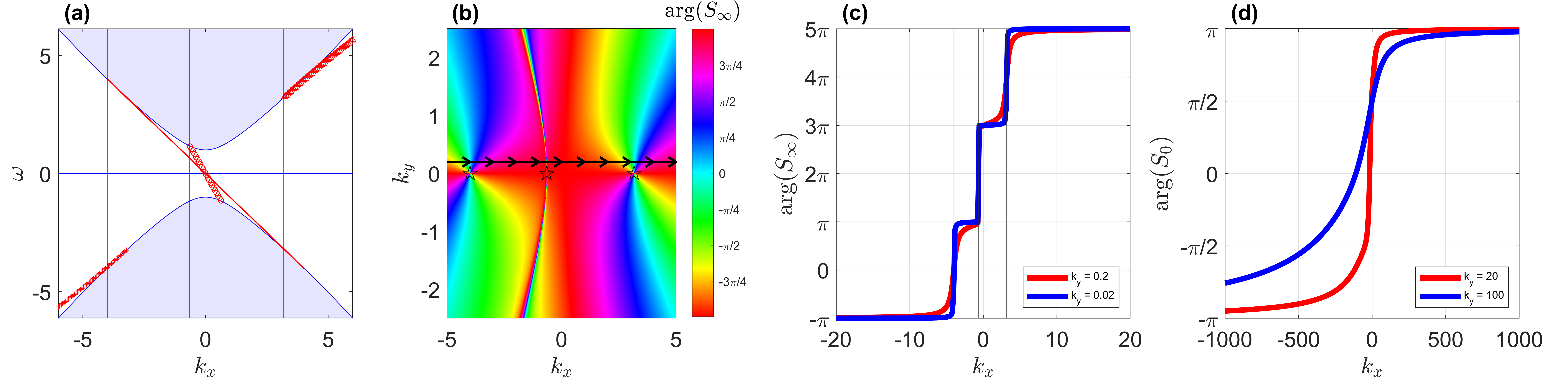}

\caption{\label{fig:adep_phase}Dispersion and scattering phase accumulation
for the parameter-dependent Hermitian boundary condition (\ref{eq:boundary_cond}).
Here $f=1,\nu=1/16,a=-1.2$. (a) Bulk and edge dispersions. Colors
are as in Fig. \ref{fig:Dispersion-multi-adep}. (b) Colormap of $\arg(S_{\infty}$)
in the $\left(k_{x},k_{y}\right)$ plane. The black arrows depict
a straight line contour along the bottom of bulk band, at constant
small $k_{y}$. Stars mark emergence of edge modes, at $K_{x,1}=-4,K_{x,2}\approx-0.64,K_{x,3}\approx3.17$,
which here correspond to singularities of $S_{\infty}$ where a root
and a pole coalesce. Note that we plot $S_{\infty}$ in the plane
for clarity, even though the definition of the scattering state (\ref{eq:psi_zeta_sp})
means it is well-defined only in the upper half-plane. (c) The phase
accumulation of the scattering matrix at finite $k_{x}$, using $S_{\infty}$,
along contours similar to the one depicted in (b), that is, a straight
line with constant $k_{y}$. The phase of $+2\pi$ is accumulated
locally at edge mode emergence, and as the constant $k_{y}$ of the
contour is decreased, the phase increase is sharper. Vertical lines
in (a),(c) indicate mode emergence. (d) The phase accumulation of
the scattering matrix for $\left|\mathbf{k}\right|\rightarrow\infty$,
using $S_{0}$, along a straight line contour with large constant
$k_{y}$, i.e., close to the top of the band. Since the direction
of increasing $k_{x}$ is inverted with relation to the circle contour
along the top of the band (which should be counter-clockwise), the
phase increase of $+2\pi$ must be interpreted as $-2\pi$, which,
together with the $3\times2\pi$ contribution of the $3$ edge modes
in (c), leads to the generalized BEC, $\left(6\pi-2\pi\right)/2\pi=2=C_{+}$.}
\end{figure*}

\subsection{Generalized BEC using scattering theory}

\subsubsection{Scattering matrix\label{subsec:hermitian-Scattering-matrix}}

The most general bulk solution in the presence of the boundary at
$y=0$ is a scattering state --- a superposition of an ingoing wave,
an outgoing wave, and an evanescent wave (bound to the edge), all
originating from the aforementioned bulk eigenmodes $\hat{\psi}^{\infty}$
{[}see Eq. (\ref{eq:psi_inf_sol_q}){]}
\begin{align}
\tilde{\psi_{s}}^{\zeta} & =\psi_{\mathrm{in}}^{\zeta}+S_{\zeta}\psi_{\mathrm{out}}^{\zeta}+T_{\zeta}\psi_{\mathrm{ev}}^{\infty}=\hat{\psi}^{\zeta}\left(k_{x},-k_{y}\right)e^{-ik_{y}y}\nonumber \\
 & +S_{\zeta}\hat{\psi}^{\zeta}\left(k_{x},k_{y}\right)e^{ik_{y}y}+T_{\zeta}\hat{\psi}^{\infty}\left(k_{x},\kev\right)e^{i\kev y},\label{eq:psi_zeta_sp}
\end{align}
where $S_{\zeta}\left(k_{x},k_{y}\right)$ is defined as the scattering
matrix (which is a scalar in the current case), determining the ratio
between the ingoing and outgoing parts. Note that this form assumes
$\re{k_{y}}>0$; indeed, for the bulk dispersion (\ref{eq:original_dispersion}),
$\frac{d\omega_{+}}{dk_{y}}\propto\text{sgn\ensuremath{\left(k_{y}\right)}}$,
so $e^{ik_{y}y}$ ($e^{-ik_{y}y}$) has positive (negative) group
velocity and therefore is outgoing (incoming) with relation to the
edge. We also define $\kappa_{\mathrm{ev}}$ as the other solution
for $k_{y}$ that admits the same frequency but leads to an evanescent
wave
\begin{equation}
\kappa_{\mathrm{ev}}\left(k_{x},k_{y}\right)=+i\sqrt{k_{y}^{2}+2k_{x}^{2}+\frac{1-2\nu f}{\nu^{2}}}\in i\ar_{+}.\label{eq:kappa_ev_def}
\end{equation}
Notice that taking the opposite sign in Eq. (\ref{eq:kappa_ev_def})
would give a divergent, hence non-physical wave, which for later use
we denote by $\kappa_{\mathrm{div}}=-\kappa_{\mathrm{ev}}$. The scattering
matrix is calculated by plugging the superposition (\ref{eq:psi_zeta_sp})
into the boundary conditions (\ref{eq:boundary_cond}). After some
algebra one gets
\begin{widetext}
\begin{align}
S_{\zeta}\left(k_{x},k_{y}\right) & =-\frac{g_{\zeta}\left(k_{x},-k_{y}\right)}{g_{\zeta}\left(k_{x},k_{y}\right)},\,\,T_{\zeta}\left(k_{x},k_{y}\right)=-\frac{h_{\zeta}\left(k_{x},k_{y}\right)}{g_{\zeta}\left(k_{x},k_{y}\right)},\nonumber \\
g_{\zeta}\left(k_{x},k_{y}\right) & =\begin{vmatrix}k_{x}\hat{u}_{\zeta}\left(k_{x},k_{y}\right)+ak_{y}\hat{v}_{\zeta}\left(k_{x},k_{y}\right) & k_{x}\hat{u}_{\infty}\left(k_{x},\kappa_{\mathrm{ev}}\right)+a\kappa_{\mathrm{ev}}\hat{v}_{\infty}\left(k_{x},\kappa_{\mathrm{ev}}\right)\\
\hat{v}_{\zeta}\left(k_{x},k_{y}\right) & \hat{v}_{\infty}\left(k_{x},\kappa_{\mathrm{ev}}\right)
\end{vmatrix},\nonumber \\
h_{\zeta}\left(k_{x},k_{y}\right) & =\begin{vmatrix}k_{x}\hat{u}_{\zeta}\left(k_{x},k_{y}\right)+ak_{y}\hat{v}_{\zeta}\left(k_{x},k_{y}\right) & k_{x}\hat{u}_{\zeta}\left(k_{x},-k_{y}\right)-ak_{y}\hat{v}_{\zeta}\left(k_{x},-k_{y}\right)\\
\hat{v}_{\zeta}\left(k_{x},k_{y}\right) & \hat{v}_{\zeta}\left(k_{x},-k_{y}\right)
\end{vmatrix}.\label{eq:gzeta_and_Szeta_def}
\end{align}
\end{widetext}

\noindent Direct inspection shows that $|S_{\zeta}|=1$, meaning it
is unitary and hence purely characterized by its phase, $\arg\left(S_{\zeta}\right)=-2\arg\left(g_{\zeta}\left(k_{x},k_{y}\right)\right)$.

We note that the solutions of the bulk (\ref{eq:psi_zeta_sp}) and
edge (\ref{eq:edge_modes_sp}), at the intersections between their
dispersions, are related in a continuous manner. We show these relations
explicitly, and note that they are independent of the boundary condition.
Simply plugging the bulk dispersion (\ref{eq:original_dispersion})
into the edge wavevectors $\k_{j}$ (\ref{eq:kappa_def}), we find
\begin{equation}
\k_{1}=ik_{y},\k_{2}=\im{\kev},\label{eq:kappa_continuity}
\end{equation}
(as $\k_{1,2}$ are interchangeable, the assignment is arbitrary).
Furthermore, this relation leads to the ratio of $u,v$ to be exactly
the same for the bulk and the edge modes at their intersection {[}see
Eq. (\ref{eq:edge_modes_sp}){]}
\begin{align}
\frac{\hat{u}_{\zeta}\left(k_{x},-k_{y}\right)}{\hat{v}_{\zeta}\left(k_{x},-k_{y}\right)} & =\lambda_{1},\nonumber \\
\frac{\hat{u}_{\infty}\left(k_{x},\kev\right)}{\hat{v}_{\infty}\left(k_{x},\kev\right)} & =\lambda_{2}.\label{eq:lambda_continuity}
\end{align}
These simple continuity relations between bulk and edge modes prove
useful in later calculations.

\subsubsection{Generalized BEC\label{subsec:Generalized-BEC}}

The BEC is the statement that the bulk index, namely, a topological
invariant, coincides with the edge index, counting the number of protected
edge modes in the gap. We briefly repeat the definitions of the bulk
index and the edge index for our system. The bulk index for the upper
band, which is the Chern number, is given in general by
\begin{multline}
C_{+}=\\
\frac{i}{2\pi}\int_{\ar^{2}}dk_{x}dk_{y}\left[\left\langle \do_{k_{x}}\psi_{+}\left|\do_{k_{y}}\psi_{+}\right.\right\rangle -\left\langle \do_{k_{y}}\psi_{+}\left|\do_{k_{x}}\psi_{+}\right.\right\rangle \right],
\end{multline}
where $\psi_{+}$ are the corresponding eigenmodes (\ref{eq:psi_inf_sol_q})
(one must note that this integral is indeed a well-defined topological
invariant in this problem only for $\nu\ne0$. For further discussion
on this point see the following references). In Refs. \citep{Tauber2018,Souslov2019}
it was shown that for the system at hand we have
\begin{equation}
C_{+}=\mathrm{sign}\left(f\right)+\mathrm{sign}\left(\nu\right).\label{eq:Chern_in_SWM}
\end{equation}
The edge index is defined by (see Refs. \citep{Hatsugai1993b,Tauber2020})
\begin{equation}
\mathcal{L}_{\mathrm{edge}}=n_{b}-n_{a},
\end{equation}
where $n_{a},n_{b}$ are the signed numbers of intersections of edge
mode dispersion curves with the bulk dispersion for the positive band
$\omega_{+}$ in the ($k_{x},\omega$) plane: $n_{b}$ (below) is
counted positive for an emerging mode from the bottom of the band,
and negative for a mode merging into the bottom of the band; and conversely
$n_{a}$ (above) is counted negative for modes emerging from the top
of the band and positive for modes disappearing into the top of the
band. In our system obviously $n_{a}=0$, because the upper bulk band
dispersion is unbounded from above, see Eq. (\ref{eq:original_dispersion}).
Thus, in our problem the BEC should have led to $C_{+}=n_{b}$. Since
Eq. (\ref{eq:Chern_in_SWM}) gives $C_{+}=2$, one would expect a
net of two edge modes to emerge from the upper bulk band for all values
of $a$. This, however, turns out not to be the case, as we now elaborate.

Numerical calculations show that under the parameter-dependent boundary
condition (\ref{eq:boundary_cond}), the number of edge modes changes
with the parameter $a$ through four regimes. (i) For $a<-\sqrt{2}$,
there are two edge modes. (ii) For $-\sqrt{2}<a<0$, there are three
edge modes. (iii) For $0<a<\sqrt{2}$, there is a single edge mode.
(iv) For $a>\sqrt{2}$ there are again two edge modes. Thus we have
a match with the Chern number for $\left|a\right|>\sqrt{2}$ and a
mismatch for $\left|a\right|<\sqrt{2}$ (see Fig. \ref{fig:Dispersion-multi-adep}).
This mismatch is accounted for by a generalization of the BEC to take
into account the behavior at $\left|\mathbf{k}\right|\to\infty$,
using a relative version of Levinson's theorem. 

The relative version of Levinson's theorem for edge modes emerging
from/merging with the bulk dispersion at finite $k_{x}$ was introduced
in Ref. \citep{Graf2012} (theorem 6.11). It states 
\begin{equation}
\lim_{k_{y}\rightarrow0^{+}}\left(\arg S_{\zeta}\left(k_{x},k_{y}\right)\right)|_{k_{x}^{1}}^{k_{x}^{2}}=2\pi n\left(k_{x}^{1},k_{x}^{2}\right),\label{eq:relative_levinson_thm}
\end{equation}
meaning that for a finite segment $\left[k_{x}^{1},k_{x}^{2}\right]$,
taking $k_{y}\rightarrow0^{+}$, i.e., approaching the bottom of the
bulk band $\omega_{+}\left(\mathbf{k}\right)$ from above, the phase
of the scattering matrix counts the net number of edge modes that
appear and disappear in that segment. Thus, for a sufficiently long
segment of the line it indeed counts all edge modes that (dis)appear
in the bottom of the upper band, namely $n_{b}$. This allows one
to arrive at a generalized BEC: The bulk Chern number $C_{+}$ can
be related to the phase accumulation of the scattering matrix $S_{\zeta}$
over an appropriate closed contour in the $(k_{x},k_{y})$ plane.
Before continuing with the consequences of this statement, it is worth
mentioning the relation of $S_{\zeta}$ to the bulk topology, which
is the at heart of the proof. The idea is that $\psi_{\mathrm{in}}=\hat{\psi}^{\zeta}\left(k_{x},-k_{y}\right),\psi_{\mathrm{out}}=\hat{\psi}^{\zeta}\left(k_{x},k_{y}\right)$
are regular sections of the sphere isomorphic to the compactified
$\left(k_{x},k_{y}\right)$ plane, $S^{2}\cong\ar^{2}\cup\left\{ \infty\right\} $,
at $S^{2}\backslash\left\{ \zeta\right\} $ and $S^{2}\backslash\left\{ \bar{\zeta}\right\} $,
respectively. Therefore, the scattering matrix acts as the transition
matrix between the sections, and its winding along an appropriate
closed circular contour on the sphere introduced in Ref. \citep{Graf2020}
is exactly the bulk Chern number $C_{+}=\frac{1}{2\pi i}\oint S_{\zeta}^{-1}dS$.
We do not repeat the entire discussion, but it is important to note
that along the chosen contour, $S_{\zeta}$ is well-defined and does
not vanish. Now, the winding of $S_{\zeta}$ along the closed contour
(which, as just said, equals $C_{+}$) can be decomposed into a $k_{y}\to0^{+}$
contribution, counting the edge modes per the relative Levinson's
theorem, and a contribution from $k_{y}\to\infty$, resulting in the
generalized BEC (see theorem 2.9 in Ref. \citep{Graf2020})
\begin{equation}
C_{+}=\mathrm{Ind}\left(S_{\zeta}\left(k_{y}\rightarrow\infty\right)\right)+n_{b},\label{eq:generalized_BEC_graf}
\end{equation}
where we define the winding number of $S_{\zeta}$ over a contour
$\gamma$ by $\mathrm{Ind}\left(S_{\zeta}\right)=\frac{1}{2\pi i}\int_{\gamma}S_{\zeta}^{-1}dS$. 

\begin{figure*}[t]
\includegraphics[scale=0.46]{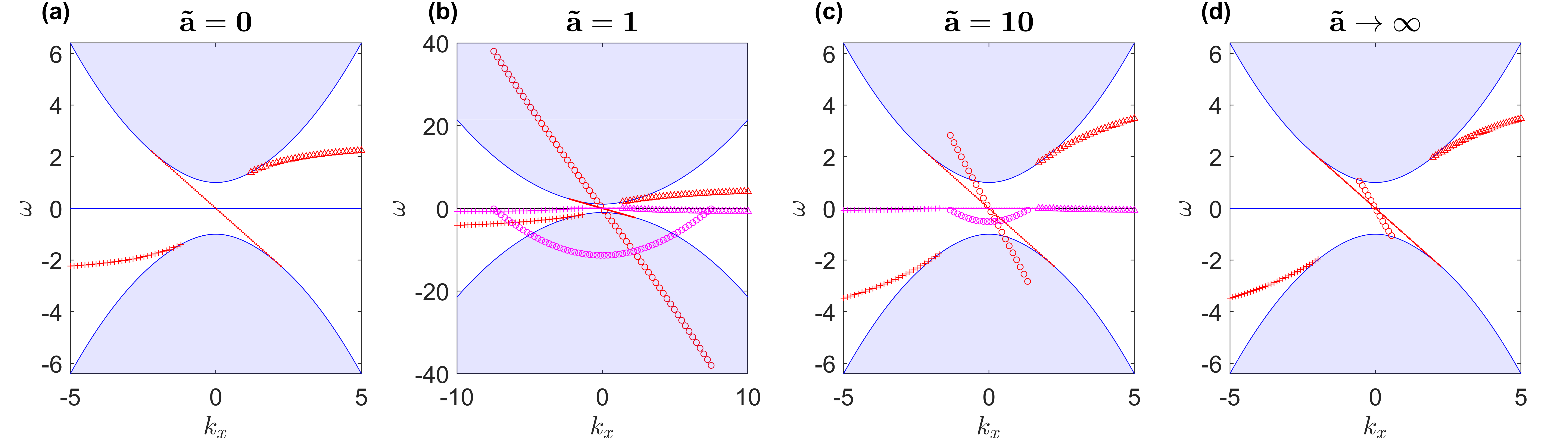}\caption{\label{fig:interpolated-Dispersion-plots}Dispersion with the non-Hermitian
boundary condition (\ref{eq:explicit_interpolated_BC}) for different
values of $\tilde{a}$. Here $f=1,\nu=0.2$. As in Fig. \ref{fig:Dispersion-multi-adep},
the real part of the bulk dispersion is in light blue/blue (its imaginary
part vanishes); the real and imaginary parts of the edge mode dispersions
are in red and pink, respectively (in panels (a),(d) only the real
part is shown, since the imaginary part vanishes). Markers of the
same shape correspond to the same edge mode. As explained in the text,
in all cases we have the $\tilde{a}$-independent mode $\omega=-k_{x}$,
which is depicted by continuous lines. (a) For $\tilde{a}=0$ (no-slip)
we have an additional edge mode emerging from the upper band (depicted
by triangles), and thus two overall. For any nonzero $\tilde{a}$
(b)--(d) we have three edge modes overall, two of which are continuously
related to the modes in the no-slip case, and the third depicted by
circles. The two $\tilde{a}$-dependent edge modes (circles and triangles)
intersect with the bulk away from its bottom, with positive $k_{y}$,
for any finite $\tilde{a}>0$ (see Fig. \ref{fig:interpolated_phase}),
since they have nonzero $\protect\im{\omega}$ and thus \textquotedblleft bypass\textquotedblright{}
the real bulk dispersion via the imaginary axis. The third edge mode
(circles), which is absent in the no-slip case, intersects the bulk
dispersion at large $k_{x}$ for small $\tilde{a}$ (it is difficult
to see in two dimensions that it does in fact intersect with the bulk
dispersion). (d) For $\tilde{a}\rightarrow\infty$ (no-stress), this
third mode dispersion becomes close to linear, $\omega\approx-2k_{x}$.}
\end{figure*}

One main idea in the proof of the relative Levinson's theorem (\ref{eq:relative_levinson_thm})
is the role of the poles of $S_{\zeta}$ in the complex $k_{y}$ plane.
We repeat it briefly for better understanding of its later generalizations.
Up to normalization we can write the scattering state (\ref{eq:psi_zeta_sp})
(suppressing $k_{x}$ dependence) as
\begin{multline}
\tilde{\psi}_{s}=g_{\zeta}\left(k_{y}\right)\hat{\psi}^{\zeta}\left(-k_{y}\right)e^{-ik_{y}y}-g_{\zeta}\left(-k_{y}\right)\hat{\psi}^{\zeta}\left(k_{y}\right)e^{ik_{y}y}\\
-h_{\zeta}\left(k_{y}\right)\hat{\psi}^{\infty}\left(\kev\right)e^{i\kev y}.
\end{multline}
Now treat $k_{x}$ as a parameter and follow the complex trajectory
of $K_{y}$, a root of $g_{\zeta}\left(k_{y}\right)$, or equally,
a pole of $S_{\zeta}\left(k_{y}\right)$. Note that as long as $\im{K_{y}}>0$,
the scattering state above is in fact an edge state, since the second
and third terms decay exponentially away from the edge, whereas for
$\im{K_{y}}<0$ the second term diverges. Thus, whenever a complex
pole $K_{y}$ of $S_{\zeta}$ approaches the real $k_{y}$ axis, an
edge mode dispersion intersects with the bulk. Note that this same
argument works for the complex roots of $S_{\zeta}$ as well, but
with the opposite imaginary part (see Fig. \ref{fig:kycomplex_Sroots}),
since $S_{\zeta}\left(k_{x},-k_{y}\right)=\left(S_{\zeta}\left(k_{x},k_{y}\right)\right)^{-1}$,
see Eq. (\ref{eq:gzeta_and_Szeta_def}). Roots of the scattering matrix
can be thought of as boundary-absorption resonances --- points in
parameter space where an incoming wave results only in a wave bound
to the edge, and no outgoing term. Since $S_{\zeta}$ is unitary for
real $k_{y}$, a complex $k_{y}$ pole of $S_{\zeta}$ can only approach
the real $k_{y}$ axis together with a corresponding root at $-k_{y}$.
Further, from the structure of $S_{\zeta}$ we conclude that the roots
and poles of the scattering matrix coincide as singularities with
finite limits, and give rise to phase of the form $\left[-\left(\bar{z}-K_{x}\right)/\left(z-K_{x}\right)\right]$
in the $\left(k_{x},k_{y}\right)$ plane, with $z=k_{x}+ik_{y}$ and
real $K_{x}$ {[}Fig. \ref{fig:adep_phase}(b){]}. This also fits
the fact that edge mode dispersion intersections with the bulk band
dispersion must occur at the bottom of the band, which is trivially
achieved at $k_{y}=0$ for any $k_{x}\in\ar$, see Eq. (\ref{eq:original_dispersion}).
Thus, the phase of the scattering matrix senses bulk-edge dispersion
intersections, by changing by $2\pi$ near them (increasing for modes
emerging from the bottom of the upper bulk band and decreasing for
modes merging into it). 

In Fig. \ref{fig:adep_phase} we demonstrate the generalized BEC numerically.
For the chosen parameter $a$, there are three edge modes in the dispersion
plot {[}Fig. \ref{fig:adep_phase}(a){]}. Accordingly, at each of
their intersections with the bulk, the phase of the scattering matrix
increases locally by $2\pi$ {[}Fig. \ref{fig:adep_phase}(c){]}.
The BEC mismatch is accounted for by counting the contribution from
the top of the band, i.e., $\left|\mathbf{k}\right|\to\infty$, and
indeed the phase of the scattering matrix accumulates $-2\pi$ there
{[}Fig. \ref{fig:adep_phase}(d){]}, namely, $\mathrm{Ind}\left(S_{\zeta}\left(k_{y}\rightarrow\infty\right)\right)=-1$.
For the numerical calculation of $n_{b}$, namely, finite-$k_{x}$
edge modes, $S_{\infty}$ is used with a straight line contour, $\gamma$,
along the bottom of the band, i.e., constant small $k_{y}$ and increasing
$k_{x}$ {[}Fig. \ref{fig:adep_phase}(b){]}. For the calculation
of $\mathrm{Ind}\left(S_{\zeta}\left(k_{y}\rightarrow\infty\right)\right)$
along the top of the band, $S_{0}$ is used, with a similar straight
line contour, $\gamma$, at large constant $k_{y}$. 

\section{Non-Hermitian edge problem\label{sec:Non-hermitian-edge-problem}}

In this section we add dissipation to the edge problem via a non-Hermitian
boundary condition, while keeping the bulk problem Hermitian. We show
that the edge modes decay in time, and accordingly the scattering
matrix is not unitary. However, a modified version of the relative
Levinson's theorem (\ref{eq:modified_levinson_thm_orr}) still holds,
with the appropriate contours shifted away from the bottom of the
upper bulk band, which maintains the generalized BEC in this case.

\subsection{Edge problem}

Consider a family of ``partial-slip'' boundary conditions \citep{J.C1879,Torre2015,Falkovich2017,Kiselev2019},
with the real parameter $\tilde{a}\geq0$, which interpolates between
the no-slip condition $u|_{y=0}=0$ at $\tilde{a}=0$, and the no-stress
condition $T_{xy}|_{y=0}=0$ as $\tilde{a}\rightarrow\infty$, where
$T_{xy}$ is the stress tensor:
\begin{equation}
v|_{y=0}=0,u|_{y=0}=-\tilde{a}T_{xy}|_{y=0}.\label{eq:interpolated_BC}
\end{equation}
The first equation means vanishing perpendicular velocity at the boundary
as before, and the second relates the velocity parallel to the edge
with the force acting on the fluid at the edge, i.e., it physically
represents friction at the edge. Thus, $\tilde{a}$ may be interpreted
as the slip length for particles at the boundary. Note that this boundary
condition is inherently dissipative, even without dissipative viscosity.
We prove this rigorously in Appendix \ref{sec:App.proof-NHBCisdissipative}.
Plugging in the expression for the stress tensor, and using the translation
invariance along the edge, the boundary conditions can be written
as
\begin{equation}
v|_{y=0}=0,u|_{y=0}=\tilde{a}\nu\left(ik_{x}u-\partial_{y}v\right)|_{y=0}.\label{eq:explicit_interpolated_BC}
\end{equation}

The bulk spectrum is not modified by the new boundary conditions.
As for the edge modes, following the same recipe as in the Hermitian
problem, we assume the ansatz (\ref{eq:edge_modes_sp}), plug it into
the boundary conditions (\ref{eq:explicit_interpolated_BC}), and
find a dispersion equation for $\omega,\k_{1},\k_{2}$ (for $u_{0},v_{0}\ne0$),
\begin{equation}
-i\tilde{a}\nu\left(\omega^{2}-k_{x}^{2}\right)=\left[-k_{x}+\omega\nu\left(\kappa_{2}+\kappa_{1}\right)\right]\left(ik_{x}\tilde{a}\nu-1\right),\label{eq:interpolated_dispersion_eq}
\end{equation}
which can be simplified to finding the roots of a complex polynomial
of degree 4 of $\omega^{2}$ (up to verifying $\re{\k_{j}}>0$), using
Eq. (\ref{eq:kappa_def})
\begin{widetext}
\begin{multline}
0=\tilde{a}^{4}\nu^{4}\omega^{8}+\left[-2\tilde{a}^{2}\nu^{2}\left(2i\tilde{a}\nu k_{x}+4\tilde{a}^{2}\nu^{2}k_{x}^{2}-\left(1-i\tilde{a}\nu k_{x}\right)^{2}\left(2k_{x}^{2}\nu^{2}+1-2f\nu\right)\right)+4\nu^{2}\left(1-i\tilde{a}\nu k_{x}\right)^{4}\right]\omega^{6}\\
+\left\{ \left[2i\tilde{a}\nu k_{x}+4\tilde{a}^{2}\nu^{2}k_{x}^{2}-\left(1-i\tilde{a}\nu k_{x}\right)^{2}\left(2k_{x}^{2}\nu^{2}+1-2f\nu\right)\right]^{2}-2\tilde{a}^{2}\nu^{2}\left(k_{x}^{2}-4i\tilde{a}\nu k_{x}^{3}-4\tilde{a}^{2}\nu^{2}k_{x}^{4}\right)\right.\\
\left.-\left(1-i\tilde{a}\nu k_{x}\right)^{4}\left[\left(2\nu^{2}k_{x}^{2}+1-2f\nu\right)^{2}+4\nu f-1\right]\right\} \omega^{4}\\
+2\left[2i\tilde{a}\nu k_{x}+4\tilde{a}^{2}\nu^{2}k_{x}^{2}-\left(1-i\tilde{a}\nu k_{x}\right)^{2}\left(2k_{x}^{2}\nu^{2}+1-2f\nu\right)\right]\left(k_{x}^{2}-4i\tilde{a}\nu k_{x}^{3}-4\tilde{a}^{2}\nu^{2}k_{x}^{4}\right)\omega^{2}+\left(k_{x}^{2}-4i\tilde{a}\nu k_{x}^{3}-4\tilde{a}^{2}\nu^{2}k_{x}^{4}\right)^{2}.\label{eq:interpolated_edge_dispersion_eq_polynomial}
\end{multline}
\end{widetext}

It is easy to check that the Kelvin wave solution $\omega=-k_{x}$,
$v=0$, still holds in this case, and is independent of $\tilde{a}$,
since $v=0$ means that the boundary condition (\ref{eq:explicit_interpolated_BC})
is reduced to $u|_{y=0}=0$. Equivalently, we see that the Kelvin
wave edge mode obeys both the no-slip condition and the no-stress
condition, hence it obeys a linear interpolation between them. Numerically,
one finds the other edge modes have complex dispersions with $\im{\omega}<0$,
for $\tilde{a}>0$, meaning that they indeed decay in time (see Fig.
\ref{fig:interpolated-Dispersion-plots}). This leads to a result
unique to the non-Hermitian case --- the edge mode dispersions can
intersect with the bulk band away from its bottom, since the edge
dispersion can ``bypass'' the bulk dispersion bottom via the imaginary
frequency axis.

We briefly describe the dependence of edge mode dispersions on the
parameter $\tilde{a}$ (see Fig. \ref{fig:interpolated-Dispersion-plots}),
with the important point being that there is such a dependence, so
the usual BEC is broken, as in the Hermitian case. For $\tilde{a}=0$
(no-slip), there are two edge modes emerging at finite $k_{x}$ and
no phase accumulation for $S_{0}$ at infinity; for $\tilde{a}=\infty$
(no-stress), there are three edge modes at finite $k_{x}$ and the
appropriate correction at infinity (as seen in Ref. \citep{Tauber2020})
--- these are the Hermitian cases. For any finite $\tilde{a}>0$,
there are three edge modes as well, where two of them emerge not from
the bottom of the band. One of them can be continuously related to
the mode in the no-slip case, and the other ``comes from infinity''
--- meaning that as $\tilde{a}$ decreases it emerges from the upper
band and disappears in the lower band at larger values of $k_{x}$.
For large $\tilde{a}$, it approaches the third mode in the no-stress
case, whose dispersion is almost linear, $\omega\approx-2k_{x}$ \footnote{In a non-Hermitian system there are in principle other possibilities,
such as a residual spectrum, or a point spectrum which covers a 2D
domain in the complex plane \citep{Eidelman2004}. By studying the
equation $(I\omega-H)\Psi=\Phi$ for arbitrary $\Phi$ one can show
these options do not occur, the underlying reason being that the corresponding
homogeneous equation always has exactly two modes which decay away
from the edge for any $\omega$ away from the bulk spectrum}.

\begin{figure*}[t]
\includegraphics[scale=0.46]{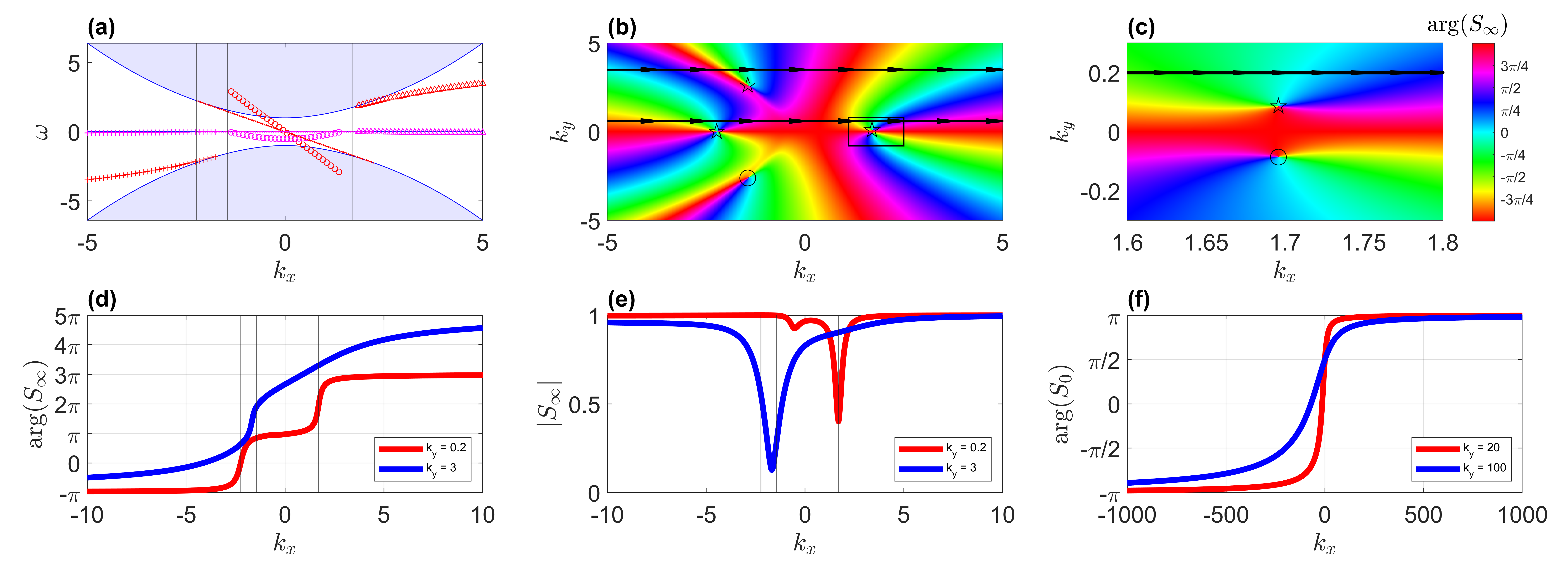}\caption{\label{fig:interpolated_phase}Dispersion and scattering phase accumulation
for the non-Hermitian boundary condition (\ref{eq:explicit_interpolated_BC}).
Here $f=1,\nu=0.2,\tilde{a}=10$. (a) Bulk and edge dispersions. Colors
are as in Fig. \ref{fig:interpolated-Dispersion-plots}. (b) Colormap
of $\arg(S_{\infty}$) in the $\left(k_{x},k_{y}\right)$ plane. Note
that we plot $S_{\infty}$ in the plane for clarity, even though the
definition of the scattering state (\ref{eq:psi_zeta_sp}) means it
is well-defined only in the upper half-plane. The edge modes emerge
for roots of $S_{\infty}$ with $K_{y}>0$, which are separated from
the poles, except for the parameter-independent mode $\omega=-k_{x}$,
which emerges at $\left(-\sqrt{f/\nu},0\right)$. These are marked
by stars, at $\left(K_{x,1},K_{y,1}\right)=\left(-\sqrt{5},0\right),\left(K_{x,2},K_{y,2}\right)\approx\left(-1.45,2.61\right),\left(K_{x,3},K_{y,3}\right)\approx\left(1.70,0.09\right)$
{[}the last one appears in panel (c){]}; the last two occur at $K_{y}>0$,
and correspond to roots of $S_{\infty}$, while the first one has
$K_{y}=0$, and thus corresponds to a singularity (coalescence of
root and pole), as in the Hermitian case. The black arrows depict
two straight line contours at constant $k_{y}$, one is above all
roots, and one is above only two of them. Circles mark poles in the
lower half-plane, at $-K_{y}$ for each root with $K_{y}>0$. (c)
Zoom in of (b) around the root and pole pair at $K_{x,3}$. We see
the separation of the root and the pole, which corresponds to the
dispersion of this mode emerging away from the bottom of the band.
The same colorbar applies to (b)--(c). (d) The phase accumulation
of the scattering matrix at finite $k_{x}$, using $S_{\infty}$,
along the straight line contours depicted in (b)--(c). We see that
only contours with $k_{y}$ larger than $K_{y}$ of the intersection
of a mode with the bulk can sense that mode, but that taking $k_{y}$
much larger than $K_{y}$ results in smeared accumulation of phase.
(e) The magnitude of the scattering matrix. It is smaller than $1$
around $k_{x}$ values where modes emerge, with the exception of the
$\omega=-k_{x}$ mode. Vertical lines in (a),(d)--(e) indicate mode
emergence (see values of $K_{x}$ above). (f) The phase accumulation
of the scattering matrix for $\left|\mathbf{k}\right|\to\infty$,
using $S_{0}$, along a straight line contour with large constant
$k_{y}$. As in the Hermitian case (see Fig. \ref{fig:adep_phase}),
since the contour direction is inverted, the phase increase of $+2\pi$
must be interpreted as $-2\pi$, which, together with the $3\times2\pi$
contribution of the $3$ edge modes in (d), leads to the generalized
BEC, $\left(6\pi-2\pi\right)/2\pi=2=C_{+}$, in the same manner as
in the Hermitian case.}
\end{figure*}

\subsection{Generalized BEC\label{subsec:interpolated-generalized-BEC}}

As before, we use the scattering state solution (\ref{eq:psi_zeta_sp})
with the boundary conditions (\ref{eq:explicit_interpolated_BC})
in order to find the scattering matrix. The structure is the same
as before
\begin{widetext}
\begin{align}
S_{\zeta}\left(k_{x},k_{y}\right) & =-\frac{\tilde{g}_{\zeta}\left(k_{x},-k_{y}\right)}{\tilde{g}_{\zeta}\left(k_{x},k_{y}\right)},\,\,T_{\zeta}\left(k_{x},k_{y}\right)=-\frac{\tilde{h}_{\zeta}\left(k_{x},k_{y}\right)}{\tilde{g}_{\zeta}\left(k_{x},k_{y}\right)},\nonumber \\
\tilde{g}_{\zeta}\left(k_{x},k_{y}\right) & =\left|\begin{pmatrix}\left(\tilde{a}\nu ik_{x}-1\right)\hat{u}_{\zeta}\left(k_{x},k_{y}\right)-\tilde{a}\nu ik_{y}\hat{v}_{\zeta}\left(k_{x},k_{y}\right) & \left(\tilde{a}\nu ik_{x}-1\right)\hat{u}_{\infty}\left(k_{x},\kappa_{ev}\right)-\tilde{a}\nu i\kappa_{ev}\hat{v}_{\infty}\left(k_{x},\kappa_{ev}\right)\\
\hat{v}_{\zeta}\left(k_{x},k_{y}\right) & \hat{v}_{\infty}\left(k_{x},\kappa_{ev}\right)
\end{pmatrix}\right|,\nonumber \\
\tilde{h}_{\zeta}\left(k_{x},k_{y}\right) & =\left|\begin{pmatrix}\left(\tilde{a}\nu ik_{x}-1\right)\hat{u}_{\zeta}\left(k_{x},k_{y}\right)-\tilde{a}\nu ik_{y}\hat{v}_{\zeta}\left(k_{x},k_{y}\right) & \left(\tilde{a}\nu ik_{x}-1\right)\hat{u}_{\infty}\left(k_{x},-k_{y}\right)+\tilde{a}\nu ik_{y}\hat{v}_{\infty}\left(k_{x},-k_{y}\right)\\
\hat{v}_{\zeta}\left(k_{x},k_{y}\right) & \hat{v}_{\infty}\left(k_{x},-k_{y}\right)
\end{pmatrix}\right|.\label{eq:interpolated_Smatrix}
\end{align}
\end{widetext}

\noindent Note that the scattering matrix is no longer unitary. We
show analytically in Appendix \ref{sec:App-detS=00003D1_interpolated_bc}
that $\left|S_{\zeta}\left(k_{x},k_{y}\right)\right|<1$ for all $k_{y}>0$,
which is the physical half-plane, where the incoming and outgoing
waves have the correct propagation direction (towards and away from
the boundary, respectively), in accordance with the dissipative nature
of the boundary condition (\ref{eq:explicit_interpolated_BC}). Conversely,
since the scattering matrix has the property $S_{\zeta}\left(k_{x},-k_{y}\right)=S_{\zeta}^{-1}\left(k_{x},k_{y}\right)$
as before, we see that $\left|S_{\zeta}\left(k_{x},k_{y}\right)\right|>1$
for all $k_{y}<0$, which is the unphysical half-plane, where the
order of incoming and outgoing waves is inverted in the scattering
state (\ref{eq:psi_zeta_sp}). Specifically, for each complex root
$K_{y}>0$, there is a corresponding pole at $-K_{y}<0$. These results
are supported by our numerical calculations, see Fig. \ref{fig:interpolated_phase}. 

We now discuss the generalized BEC in the non-Hermitian case. First
we review the numerical results shown in Fig. \ref{fig:interpolated_phase},
and then use them to motivate our more general claims. We see in the
plot of the scattering matrix phase {[}panels \ref{fig:interpolated_phase}(b)--(c){]}
that the roots and the poles of the scattering matrix are separated,
unlike the Hermitian case where they coalesce at $K_{y}=0$. From
Fig. \ref{fig:kycomplex_Sroots}, we see that all the roots are in
the physical half-plane, $K_{y}>0$; this is natural since $|S_{\zeta}(k_{x},k_{y})|<1$
only in that half-plane. However, we still find that the phase accumulated
on a contour encircling a root-pole pair is $4\pi$ as before (compare
with Fig. \ref{fig:adep_phase}). Hence, here the root-pole pair has
phase of the form of the complex function $\left\{ -\left[\bar{z}-\left(K_{x}+iK_{y}\right)\right]/\left[z-\left(K_{x}+iK_{y}\right)\right]\right\} $,
where $z=k_{x}+ik_{y}$, with non-real $K_{x}+iK_{y}$, instead of
real $K_{x}$ in the Hermitian case. In accordance with these observations,
we see {[}Fig. \ref{fig:interpolated_phase}(d){]} that in order for
the phase of the scattering matrix to count an edge mode merging/emerging
at $K_{y}$, we must take a contour ``above'' it, i.e., with $k_{y}>K_{y}$.
Indeed, the contour with the smaller $k_{y}$ does not sense the edge
mode with $K_{y}>k_{y}$. 

We thus see that the complex dispersion of the edge modes allows intersections
with the bulk band away from its bottom. This calls for a way to identify
bulk-edge dispersion intersections. For that we will show that the
above observations are in fact a special case of a general rule, namely
that the bulk-edge dispersion intersections correspond to the roots
of the scattering matrix: We use the continuity relations in Eqs.
(\ref{eq:kappa_continuity}) and (\ref{eq:lambda_continuity}) to
write the edge mode ansatz (\ref{eq:edge_modes_sp}) in terms of the
bulk eigenmodes, at the mode emergence. Thus, the requirement that
this solution obeys the boundary condition (\ref{eq:explicit_interpolated_BC})
results in the equation for the edge dispersion, at the points in
the wavevector plane where it coincides with the bulk dispersion (suppressing
$k_{x}$ dependence),
\begin{multline}
\tilde{a}\nu\left(ik_{y}-\im{\kev}\right)\hat{v}_{\infty}\left(\kev\right)\hat{v}_{\zeta}\left(-k_{y}\right)\\
=\left(ik_{x}\tilde{a}\nu-1\right)\left[\hat{u}_{\infty}\left(\kev\right)\hat{v}_{\zeta}\left(-k_{y}\right)-\hat{u}_{\zeta}\left(-k_{y}\right)\hat{v}_{\infty}\left(\kev\right)\right].
\end{multline}
Since the frequency is already set in terms of the other quantities,
this is indeed an equation in $k_{x},k_{y}$. On the other hand, inspection
of Eq. (\ref{eq:interpolated_Smatrix}) yields exactly the same equation
upon requiring $S_{\zeta}\left(k_{x},k_{y}\right)=0$.

Next, we deduce from the above observations that a modification of
the relative Levinson's theorem is required in order for the non-Hermitian
generalized BEC to hold. Since the edge mode dispersions intersect
with the bulk band at the roots of the scattering matrix, which are
separated from its poles, and are thus away from the bottom of the
band, we must consider contours with $k_{y}$ large enough. Therefore,
in order to count all edge modes, we must take a contour with $k_{y}>\tilde{K}_{y}=\max_{\a}\left(K_{y,\a}\right)$,
but this comes at a price, that the phase accumulated for modes merging/emerging
for all other edge modes (at $K_{y,\a}<\tilde{K}_{y}$) is smeared
{[}see Fig. \ref{fig:interpolated_phase}(d){]}. For this reason,
the straight line contour must be taken with its edges $k_{x_{1}}\rightarrow-\infty,k_{x_{2}}\rightarrow\infty$.
In practice, for numerical calculations it is enough to use finite
$k_{x_{1}},k_{x_{2}}$ which are large enough such that $K_{x,\a}$
are all inside of $\left(k_{x_{1}},k_{x_{2}}\right)$; For any given
$k_{y}>\tilde{K}_{y}$ and any desired resolution $\delta$, there
exist $k_{x_{1}},k_{x_{2}}$ large enough so that $\arg\left[S_{\zeta}\left(k_{x},k_{y}\right)\right]|_{k_{x_{1}}}^{k_{x_{2}}}=2\pi n-\delta$.
We note, similarly to the discussion in Ref. \citep{Graf2020} Sec.
3.1, that even though the scattering matrix has real roots embedded
in the bulk band, it retains its role as the transition matrix between
sections, since the contours taken in the $\left(k_{x},k_{y}\right)$
plane avoid these roots, as well as the singularities at $\zeta$.

We finally arrive at the proof of the non-Hermitian relative Levinson's
theorem (see Sec. \ref{par:Modified-relative-Levinson's-thm}). We
use the continuous transition between the Hermitian and non-Hermitian
regimes, given by the partial-slip boundary condition (\ref{eq:explicit_interpolated_BC}),
as well as the usual argument regarding topological invariants, that
the winding of $S_{\zeta}$, which is an integer, is by continuity
constant as a parameter is changed, as long as no roots or singularities
of $S_{\zeta}$ cross the contour. We define $b=1/\tilde{a}$, and
start from the no-stress condition at \textbf{$b=0$}. We then continuously
increase $b$ to some value $b_{0}>0$. We denote by $K_{x}\left(b\right),K_{y}\left(b\right)$
a real root of the scattering matrix $S_{\zeta}\left(k_{x},k_{y}\right)$
for the parameter $b$ of the boundary condition. The continuous change
in $b$ causes $K_{y}\left(b\right)$ to continuously increase from
$0$ to $K_{y}\left(b_{0}\right)$, and $K_{x}\left(b\right)$ may
move as well. Along this transition, for any value of $b$ we take
a straight line contour along $k_{x}\in\left(-\infty,\infty\right)$,
and constant $k_{y}\left(b\right)=K_{y}\left(b\right)+\epsilon$.
For this choice, we see that for any $b\in\left[0,b_{0}\right]$,
the phase of the scattering matrix along the contour counts this edge
mode correctly, since the winding of $S_{\zeta}$ is an integer and
a continuous function of $k_{x},k_{y},b$, and the chosen contour
avoids the root $K_{y}\left(b\right)$ of $S_{\zeta}$, as well as
its singularity $\zeta$. Thus, we can use a single contour so that
the phase of the scattering matrix counts this edge mode for all $b\in\left[0,b_{0}\right]$
by choosing $k_{y}=K_{y}\left(b_{0}\right)+\epsilon$. Indeed, still
this single contour is a continuous change from the one taken in the
Hermitian case for $b=0$, and thus the winding of $S_{\zeta}$ along
the two of them is the same. To complete the proof of the non-Hermitian
generalized BEC we note that the same argument is valid for the winding
of the scattering matrix along the top of the band, $\mathrm{Ind}\left(S_{\zeta}\left(k_{y}\rightarrow\infty\right)\right)$.
We again start from the no-stress condition at $b=0$ and increase
$b$ continuously. There is a straight line contour at large constant
$k_{y}$ such that $\mathrm{Ind}\left(S_{\zeta}\left(k_{y}\rightarrow\infty\right)\right)=-1$
(see Fig. \ref{fig:adep_phase}), which remains constant under the
above transition, since it is a continuous function of $b$. Thus
the sum $\mathrm{Ind}(S_{\zeta}(k_{y}\to\infty))+n_{b}$ remains equal
to its value at $b=0$, that is, to the bulk Chern number $C_{+}$,
which is the non-Hermitian generalized BEC, Eq. (\ref{eq:NH_generalized_BEC}).

A final point relates to the nature of the edge modes at infinity.
In Ref. \citep{Graf2020} it was shown that even when there is no
mismatch in the regular BEC, there is an anomaly in Levinson's theorem
at infinity, for the boundary conditions (\ref{eq:boundary_cond}).
In Appendix \ref{sec:APP-Analytic-continuation} we show a different
result for the partial-slip condition (\ref{eq:explicit_interpolated_BC}):
While the anomaly appears for any $\tilde{a}>0$, it vanishes for
the no-slip condition, $\tilde{a}=0$.

\section{Large odd viscosity\label{sec:large-odd-viscosity}}

Here we consider the case of large values of odd viscosity, $f\nu>1/2$
(and comment on the regime $1/4\leq f\nu\leq1/2$). In this case we
will see that the minimum of the band dispersion is no longer a single
point but rather a circle in the $\left(k_{x},k_{y}\right)$ plane,
which will require a modification in the contour we use. More importantly,
we will find a region in which there are two propagating bulk solutions
at given $\omega,k_{x}$, hence we will get a $2\times2$ scattering
matrix $S$, rather than a scalar. First we review the properties
of the problem in this regime, which apply to both the Hermitian and
non-Hermitian cases, and then we treat each system separately and
demonstrate the generalized BEC numerically. We stress that even though
the shape of the bulk dispersion is changed, its topology remains
as before, namely, by Eq. (\ref{eq:Chern_in_SWM}), we have $C_{+}=2$.

We assume $f,\nu>0$ throughout, and consider first the shape of the
bulk dispersion. From Eq. (\ref{eq:original_dispersion}), for $f\nu>1/2$
we see that $\omega_{+}\left(\mathbf{k}\right)$ assumes the shape
of a Mexican hat, instead of a parabola in the previous case. We note
that for $1/4\le f\nu\le1/2$ some calculations are more cumbersome
{[}it is not obvious that $\kappa_{i}\in\mathbb{R}$, see Eq. (\ref{eq:kappa_def}){]}
but the results are the same as for $f\nu<1/4$, as we will briefly
explain below. Therefore, for the rest of this section we discuss
the regime $f\nu>1/2$. The minimum of the Mexican hat $\omega_{+}\left(k_{x},k_{y}\right)$
dispersion is then achieved for every point $\left(k_{x},k_{y}\right)$
on the circle (see Ref. \citep{Souslov2019} for a similar discussion),
\begin{equation}
k_{x}^{2}+k_{y}^{2}=\frac{2f\nu-1}{2\nu^{2}}\equiv K_{c}^{2}.\label{eq:bigfnu_circle}
\end{equation}

The second important effect comes from the definitions of $\k,\kev$
in Eqs. (\ref{eq:kappa_def}) and (\ref{eq:kappa_ev_def}), for the
edge modes and bulk scattering problem, respectively. It turns out
that inside the circle (\ref{eq:bigfnu_circle}), the inner root in
Eq. (\ref{eq:kappa_def}) is complex, making the solutions $\kappa_{j}$
complex. This appears not to be a problem, since in order for the
states in the ansatz (\ref{eq:uv_bound_form}) to be bound it is sufficient
to require $\re{\kappa_{j}}>0$, regardless of the imaginary part.
Because every complex number has two square roots which are additive
inverses, we can always choose one of them with $\re{\kappa_{j}}>0$,
leaving previous calculations of the edge modes intact. For $\kappa_{\mathrm{ev}}$,
appearing in the bulk mode scattering problem, there is a more significant
modification. From Eq. (\ref{eq:kappa_ev_def}) we find that $\kappa_{\mathrm{ev}}\in i\mathbb{R}$
if and only if
\begin{equation}
2k_{x}^{2}+k_{y}^{2}>\frac{2\nu f-1}{\nu^{2}},
\end{equation}
which was trivially satisfied for $f\nu<1/2$; as a result, the behavior
for $1/4\leq f\nu\leq1/2$ is similar to $f\nu<1/4$, as stated above.
In contrast, for $f\nu>1/2$ there is an elliptical domain in $\left(k_{x},k_{y}\right)$,
\begin{equation}
2k_{x}^{2}+k_{y}^{2}\leq\frac{2\nu f-1}{\nu^{2}}\equiv K_{e}^{2},\label{eq:bigfnu_ellipse}
\end{equation}
inside which $\kappa_{\mathrm{ev,div}}$ are real and opposite in
sign, i.e., they do not correspond to evanescent/divergent solutions
anymore, but rather to propagating waves. Hence, when constructing
the superposition state (\ref{eq:psi_zeta_sp}), there are four terms:
two incoming and two outgoing waves. This requires revising the calculation
of the scattering matrix, which is now of dimension $2\times2$, and
generalizing the definition of its phase. Note that the circle (\ref{eq:bigfnu_circle})
is contained inside the ellipse, with tangent points at $k_{x}=\pm K_{c},k_{y}=0$.

Let us outline the calculation of the scattering matrix in this case.
For clarity we denote in this section $k_{y,1}=k_{y},k_{y,2}=-\kappa_{\mathrm{ev}}\left(k_{y}\right)>0$,
and recall all terms in the scattering state are multiplied by $e^{i(k_{x}x-\omega t)}$.
Note that the circle (\ref{eq:bigfnu_circle}) is the solution to
the equation $\left|k_{y,2}\left(k_{y,1}\right)\right|=\left|k_{y,1}\right|$
{[}see Eq. (\ref{eq:kappa_ev_def}){]}, and for $k_{y,1}$ outside
the circle we find $k_{y,2}\left(k_{y,1}\right)$ inside, and vice
versa. We choose a convention for the scattering state based on the
case where $k_{y,1}$ is positive and outside the circle (but still
inside the ellipse), so as before $e^{-ik_{y,1}y}$ is an incoming
wave. Here it is important to note that the dependence of the group
velocity on $k_{y}$ is inverted inside the circle (\ref{eq:bigfnu_circle}),
so $e^{ik_{y,2}y}$ is an incoming wave for the corresponding $k_{y,2}>0$
inside the circle. Accordingly, we define the full scattering state,
for $k_{y,1},k_{y,2}$ in their respective domains, by
\begin{multline}
\tilde{\psi_{s}}=A\hat{\psi}^{\infty}\left(k_{x},-k_{y,1}\right)e^{-ik_{y,1}y}+B\hat{\psi}^{\infty}\left(k_{x},k_{y,2}\right)e^{ik_{y,2}y}\\
+C\hat{\psi}^{\infty}\left(k_{x},k_{y,1}\right)e^{ik_{y,1}y}+D\hat{\psi}^{\infty}\left(k_{x},-k_{y,2}\right)e^{-ik_{y,2}y},\label{eq:bigfnu_scattering_state}
\end{multline}
where \emph{the first two terms are incoming waves}. The scattering
matrix $S$ is defined by the linear relation between the coefficients
of outgoing and incoming terms
\begin{equation}
\begin{pmatrix}C\\
D
\end{pmatrix}=S\begin{pmatrix}A\\
B
\end{pmatrix}.
\end{equation}
As usual in scattering problems, by linearity, the simplest way to
calculate the matrix elements is to consider each incoming wave separately,
and to use the boundary conditions to find how it affects the amplitudes
of the two outgoing waves. As the $2\times2$ matrix form of $S$
is only relevant inside the ellipse, it is sufficient to consider
$\hat{\psi}^{\infty}$ {[}see Eq. (\ref{eq:psi_inf_sol_q}){]}, without
moving the singularity from infinity. Another conclusion is that one
must count separately the edge modes inside the ellipse and outside
of it by ``gluing'' the contours of each domain at the domain edges
(see below). 

\begin{figure*}[t]
\includegraphics[scale=0.48]{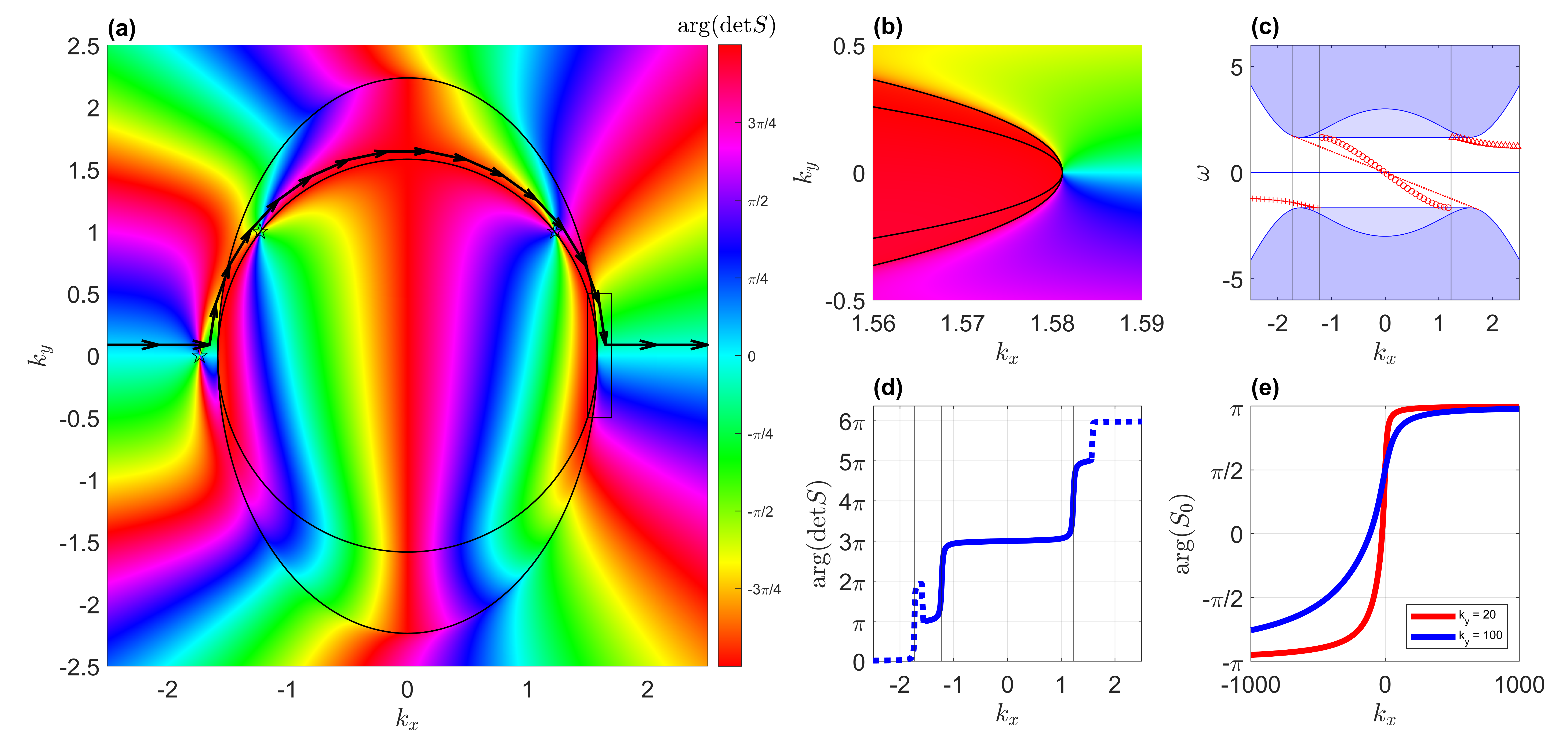}\caption{\label{fig:bigfnu_dispersion_args}Dispersion and scattering phase
accumulation inside the ellipse (\ref{eq:bigfnu_ellipse}), for large
$f\nu$, with the parameter-dependent Hermitian boundary condition
(\ref{eq:boundary_cond}). Here $f=3,\nu=1,a=-1.2$. (a) Colormap
of $\arg\left(\det S\right)$ in the $\left(k_{x},k_{y}\right)$ plane.
Inside the ellipse $S$ is a $2\times2$ matrix, and outside of it
$S_{\infty}$ is a scalar. They are continuously related up to sign
correction {[}i.e., the plot depicts $-S_{\infty}$ for clarity, see
Eq. (\ref{eq:bigfnu_Scontinuity_ellipse}){]} and the singularities
at $k_{x}=\pm K_{c},k_{y}=0$, where both are ill-defined. The stars
mark emergence of edge modes, at $\left(K_{x,1},K_{y,1}\right)=\left(-\sqrt{3},0\right),\left(K_{x,j},K_{y,j}\right)\approx\left(\pm1.23,0.998\right),j=2,3$,
which here correspond to singularities of $\det S$ where a root and
a pole coalesce. The black arrows represent the contour used in (d):
Straight lines along $k_{y}\approx0.03$ outside the ellipse, glued
to an $\approx88^{\circ}$ arc along the upper half-circle with radius
$\approx1.01\cdot K_{c}$. (b) Zoom in on (a) around the right end
of the ellipse. The discontinuity is clearly seen here to occur only
at the singularity point. The same colorbar applies to (a)--(b).
(c) Bulk and edge dispersions. Colors are as in Fig. \ref{fig:adep_phase}.
The horizontal blue lines and light-blue shaded areas indicate bulk
dispersion projection onto the plane $\left(k_{x},\omega\right)$
from the three-dimensional Mexican hat shape, filling the convex hull.
We see three edge mode dispersions emerging from the bulk dispersion.
One is the parameter-independent $\omega=-k_{x}$ emerging outside
the ellipse, and the two other modes emerge inside the ellipse, with
$K_{y}\protect\ne0$. (d) The phase accumulation of $\det S$ along
the contour depicted in (a). Dashed lines represent the phase of the
scalar $-S_{\infty}$ outside the ellipse, and continuous lines represent
the phase of $\det S$ inside. The phase increases by $2\pi$ for
each mode emergence in a localized manner along this contour. The
vertical lines in (c)--(d) indicate the $k_{x}$ values at which
the edge modes emerge (see above). Additional phase changes of $\pm\pi$
are seen at the edges of the ellipse. These artifacts of the gluing
between regimes cancel out and do not affect the BEC. (e) The phase
accumulation of the scattering matrix for $|\mathbf{k}|\to\infty$,
using $S_{0}$, along a straight line contour with large constant
$k_{y}$. As before (see Fig. \ref{fig:adep_phase}), since the contour
direction is inverted, the phase increase of $+2\pi$ must be interpreted
as $-2\pi$, which, together with the $3\times2\pi$ contribution
of the $3$ edge modes in (d), leads to the generalized BEC, $\left(6\pi-2\pi\right)/2\pi=2=C_{+}$.}
\end{figure*}

We note that for the above choice $k_{y,1},k_{y,2}$ inside the ellipse
are (trivially) continuously related to $k_{y},\kev$ outside, respectively,
with both $k_{y,2},\kev$ vanishing along the ellipse {[}see Eq. (\ref{eq:kappa_ev_def}){]}.
Thus we see that the terms in the first column of $S$ inside the
ellipse, namely the terms $S_{11},S_{21}$, have the same meaning
as $S_{\infty},T_{\infty}$ outside {[}see Eq. (\ref{eq:gzeta_and_Szeta_def}){]},
that is, they relate the amplitude of the incoming wave with those
of the outgoing and evanescent waves, respectively. Indeed, taking
the limits $k_{y,2}\rightarrow0$ inside and $\kev\rightarrow0$ outside
leads to
\begin{equation}
S=\begin{pmatrix}S_{\infty} & 0\\
T_{\infty} & -1
\end{pmatrix}\label{eq:bigfnu_Scontinuity_ellipse}
\end{equation}
along the ellipse {[}see Eqs. (\ref{eq:bigfnu_Smatrix}),(\ref{eq:GforS_interpolated_bigfnu})
below for the Hermitian and non-Hermitian cases, respectively{]}.
As a result, we see that $S_{\infty}$ outside the ellipse is continuously
related to $\det S$ (whose role will be explained below) inside,
up to a minus sign. As in Sec. \ref{subsec:hermitian-Scattering-matrix}
we can again see continuity relations at the bulk-edge dispersion
intersection between $k_{y,1},k_{y,2}$ of the bulk scattering state
and $\k_{1},\k_{2}$ of the edge state ($j=1,2$ and are interchangeable),
\begin{equation}
\k_{j}=ik_{y,j}.\label{eq:bigfnu_continuity_kykappa}
\end{equation}

We stress that the above form of the scattering wave (\ref{eq:bigfnu_scattering_state})
is not unique, with the particular convention specified above taken
for the sake of continuity. A similar choice was made in Sec. \ref{subsec:hermitian-Scattering-matrix},
where we set $e^{-ik_{y}y}$ as the incoming term in Eq. (\ref{eq:psi_zeta_sp}),
which fixed $k_{y}>0$ as the physical half-plane, and $k_{y}<0$
as unphysical, namely, that in the unphysical half-plane the scattering
matrix relates the incoming waves as dependent on the outgoing waves.
This claim is equivalently represented by the property $S_{\infty}\left(-k_{y}\right)=\left(S_{\infty}\left(k_{y}\right)\right)^{-1}$
that we saw above. Similarly, in the current case we conclude that
for the given choice in Eq. (\ref{eq:bigfnu_scattering_state}), only
the region between the ellipse and the circle with $k_{y}>0$ corresponds
to the physical scattering matrix, whereas the other domains inside
the ellipse are all unphysical due to the signs of $k_{y,1},k_{y,2}$,
meaning that the outgoing wave precedes the incoming one in at least
one of the terms: (i) Positive $k_{y,1}$ inside the circle makes
both $e^{-ik_{y,1}y}$ and $e^{ik_{y,2}y}$ outgoing, since $k_{y,2}$
is positive and outside the circle. (ii) Negative $k_{y,1}$ inside
the circle makes $e^{ik_{y,2}y}$ outgoing, again since $k_{y,2}$
is positive and outside the circle. (iii) Negative $k_{y,1}$ outside
the circle makes $e^{-ik_{y,1}y}$ outgoing. In particular, we get
that case (i) above is related to the inverse scattering matrix $S^{-1}$,
since all ingoing and outgoing terms are inverted. Indeed, for both
Hermitian and non-Hermitian cases we find the $2\times2$ scattering
matrix satisfies $S\left(-k_{y,1},-k_{y,2}\right)=\left(S\left(k_{y,1},k_{y,2}\right)\right)^{-1}$
{[}see Eqs. (\ref{eq:bigfnu_Smatrix}),(\ref{eq:GforS_interpolated_bigfnu})
below{]}. Note that we could have chosen other conventions for the
scattering state, for example, to define negative $k_{y,1}$ inside
the circle and positive $k_{y,2}$ outside the circle, such that the
incoming wave terms would be $e^{-ik_{y,1}y},e^{-ik_{y,2}y}$. This
would of course change the expression for $S$ (and the contours taken
below), but not the underlying topological structure.

Returning to our convention (\ref{eq:bigfnu_scattering_state}), the
inversion of $S$ when $k_{y,1}$ moves from outside the circle to
its inside, is related to the fact that $S$ is defective along the
circle. For our scattering state (\ref{eq:bigfnu_scattering_state})
we see that both $k_{y,1}$ and $k_{y,2}$ are positive, so that along
the edge of the circle we have $k_{y,2}\left(k_{y,1}\right)=k_{y,1}$
(see above), i.e., the two pairs of incoming and outgoing terms coalesce.
As seen in the above discussion, the edge of the circle separates
the physical and unphysical domains, as did the $k_{x}$ axis in the
small odd viscosity regime. Hence, even though the $2\times2$ scattering
matrix is not well-defined along the edge of the circle, we can still
discuss its formal properties there. One such property is $\det S=-1$
along the edge of the circle {[}see Eqs. (\ref{eq:bigfnu_Smatrix}),(\ref{eq:GforS_interpolated_bigfnu}){]},
in the same way that $S_{\infty}\left(k_{x},k_{y}=0\right)=-1$ in
the small viscosity regime. We also conclude that for both $S_{\infty}$
outside the ellipse and $\det S$ inside, we have singularities at
the edges of the ellipse, $k_{x}=\pm K_{c},k_{y}=0$, since $k_{y,1}=k_{y,2}=0$
inside and similarly $k_{y}=\kev=0$ outside, so the scattering state
is ill-defined.

Regarding the calculation of the phase of $S$ for the relative Levinson's
theorem, the correct expression for the $2\times2$ scattering matrix
is $\arg\left(\det S\right)$, as hinted in Ref. \citep{Graf2012}
in Sec. 6.3. There, in the proof of the BEC, the scattering matrix
plays the role of the transition matrix, and thus its winding is the
first Chern number. For the transition matrix it is known that the
winding is given by the winding of the product of its eigenvalues,
namely its determinant. In Appendix \ref{sec:App-Equivalence-of-numerical-methods}
we discuss two methods for numerical evaluation of the phase of the
scattering matrix, and prove their equivalence, from which the utilization
of $\det S$ is also evident. 

\subsection{Hermitian edge problem\label{subsec:bigfnuHermitian-case}}

The results of the calculation in the Hermitian case, with the parameter
dependent boundary condition (\ref{eq:boundary_cond}), are summarized
by
\begin{widetext}
\begin{align}
S\left(k_{x},k_{y,1}\right) & =-\frac{1}{G\left(k_{x},k_{y,1},-k_{y,2}\right)}\begin{pmatrix}G\left(k_{x},-k_{y,1},-k_{y,2}\right) & G\left(k_{x},k_{y,2},-k_{y,2}\right)\\
G\left(k_{x},k_{y,1},-k_{y,1}\right) & G\left(k_{x},k_{y,1},k_{y,2}\right)
\end{pmatrix},\nonumber \\
G\left(k_{x},k_{y,1},k_{y,2}\right) & =\begin{vmatrix}k_{x}\hat{u}_{\infty}\left(k_{x},k_{y,1}\right)+ak_{y,1}\hat{v}_{\infty}\left(k_{x},k_{y,1}\right) & k_{x}\hat{u}_{\infty}\left(k_{x},k_{y,2}\right)+ak_{y,2}\hat{v}_{\infty}\left(k_{x},k_{y,2}\right)\\
\hat{v}_{\infty}\left(k_{x},k_{y,1}\right) & \hat{v}_{\infty}\left(k_{x},k_{y,2}\right)
\end{vmatrix},\label{eq:bigfnu_Smatrix}
\end{align}
\end{widetext}

\noindent The following properties are immediately found upon inspection
\begin{align}
G\left(k_{x},k_{y,1},k_{y,2}\right) & =-G\left(k_{x},k_{y,2},k_{y,1}\right),\nonumber \\
G\left(k_{x},-k_{y,1},-k_{y,2}\right) & =-\overline{G\left(k_{x},k_{y,1},k_{y,2}\right)},\label{eq:bigfnu_hermitian_Gproperties}
\end{align}
which together imply the numerators in the off-diagonal terms in $S$
are real. We thus find
\begin{equation}
\det S=\frac{G\left(k_{x},-k_{y,1},k_{y,2}\right)}{G\left(k_{x},k_{y,1},-k_{y,2}\right)},\label{eq:bigfnu_hermitian_detS}
\end{equation}
from which, using the above properties, we see that $\left|\det S\right|=1$,
in consistency with the unitarity of $S$ \footnote{To be more precise, $S$ as written above is not unitary, but similar
to a unitary matrix $\tilde{S}$ by $\tilde{S}=TST^{-1}$, where $T=\text{diag\ensuremath{\left(1,\sqrt{\left|k_{y,2}/k_{y,1}\right|}\right)}}$.
This is due to the dependence of the amplitude of the eigenmodes (\ref{eq:psi_inf_sol_q})
on the wavevector $\mathbf{k}$}.

Similarly to the case of small odd viscosity with a Hermitian boundary
condition, numerical calculation shows the edge mode dispersions intersect
with the bulk at the bottom of the band, which here is the circle
(\ref{eq:bigfnu_circle}), with finite $K_{y}$. Thus, we split the
contour to two parts which we glue together: (i) Outside and away
from the ellipse we keep the straight line contour along the bottom
of the band as before, see Eq. (\ref{eq:original_dispersion}). (ii)
Inside the ellipse and close to it outside we take the upper half-circle
in the clockwise direction, namely $k_{x}^{2}+k_{y}^{2}=K_{c}^{2}+\epsilon$,
with positive $k_{y}$ and small $\epsilon$. Our numerical results
indeed show the phase of $\det S$ accumulated along this path inside
the ellipse counts the edge modes according to the relative Levinson's
theorem, see Fig. \ref{fig:bigfnu_dispersion_args}. Furthermore,
the proof of the standard (Hermitian) relative Levinson's theorem
follows the same route as in Sec. \ref{subsec:Generalized-BEC}, where
we treat $k_{x}$ as a parameter and follow a complex root $K_{y}$
of $S$ as it approaches the real $k_{y}$ axis. The point to note
here is that this argument works for the scattering state with two
incoming and two outgoing terms (\ref{eq:bigfnu_scattering_state}),
since both $k_{y,2},k_{y,1}$ approach the real $k_{y}$ axis together,
because of the dependence $k_{y,2}\left(k_{y,1}\right)$ {[}see Eq.
(\ref{eq:kappa_ev_def}){]}.

To summarize, even though this case requires more cumbersome calculations
for the $2\times2$ scattering matrix, by using $\arg\left(\det S\right)$
instead of $\arg\left(S\right)$, and a different contour inside the
ellipse, the topological structure of the original problem is kept:
The number of edge modes changes with the parameter $a$ through four
regions as in Sec. \ref{subsec:Generalized-BEC}, and corrected at
infinity in the same manner, so that the generalized BEC (\ref{eq:generalized_BEC_graf})
holds.

\begin{figure*}[t]
\includegraphics[scale=0.47]{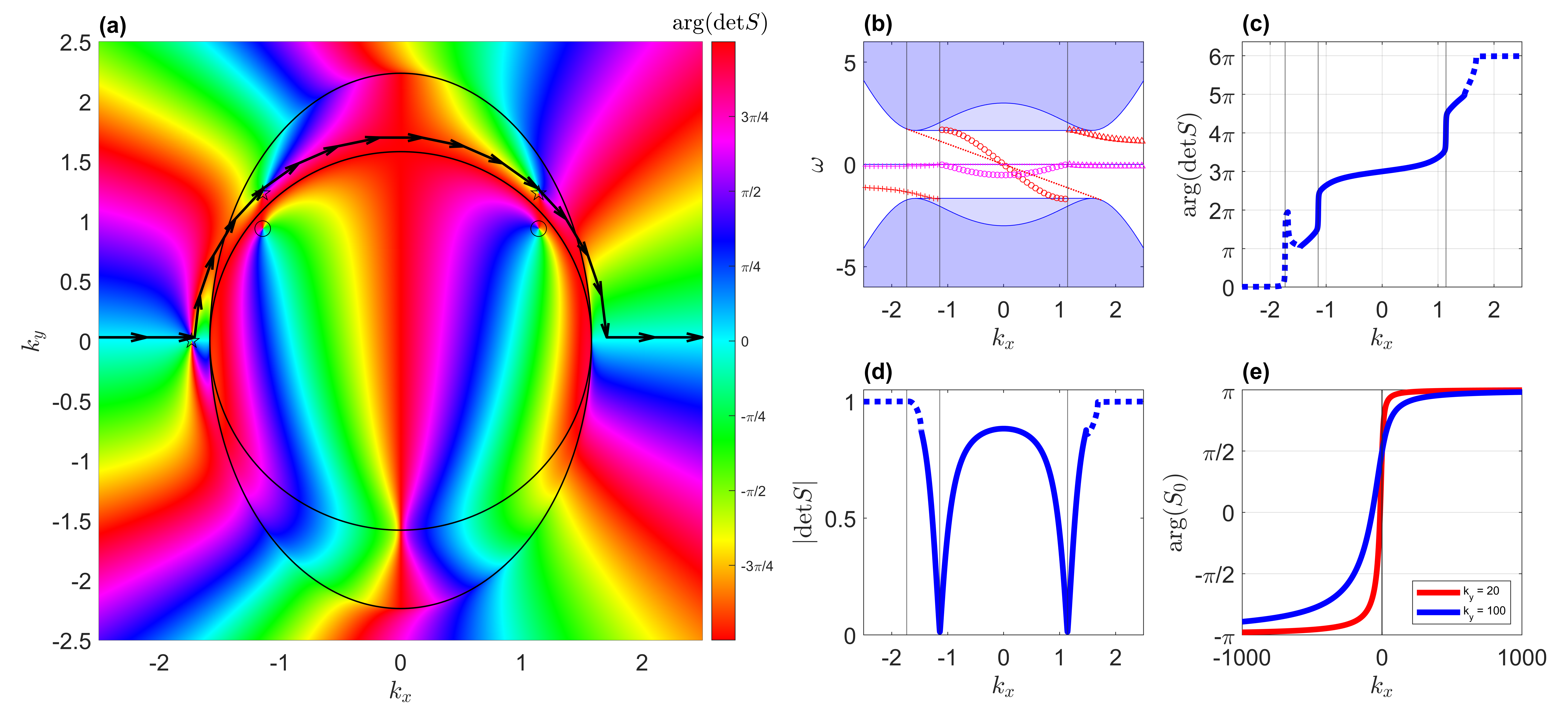}\caption{\label{fig:interpolated_bigfnu_dispersion_args}Dispersion and scattering
phase accumulation inside the ellipse (\ref{eq:bigfnu_ellipse}),
for large $f\nu$, with the non-Hermitian boundary condition (\ref{eq:explicit_interpolated_BC}).
Here $f=3,\nu=1,\tilde{a}=2$. (a) Colormap of $\arg\left(\det S\right)$
in the $\left(k_{x},k_{y}\right)$ plane. Inside the ellipse $S$
is a $2\times2$ matrix, and outside of it $S_{\infty}$ is a scalar.
They are continuously related up to sign correction {[}i.e., the plot
depicts $-S_{\infty}$ for clarity, see Eq. (\ref{eq:bigfnu_Scontinuity_ellipse}){]},
and the singularities at $k_{x}=\pm K_{c},k_{y}=0$, where both are
ill-defined. The stars mark emergence of edge modes, at $\left(K_{x,1},K_{y,1}\right)=\left(-\sqrt{3},0\right),\left(K_{x,j},K_{y,j}\right)\approx\left(\pm1.14,1.23\right),j=2,3$,
which inside the ellipse correspond to roots of $\det S$. The circles
mark poles of $\det S$. Note that the roots are outside the circle
(\ref{eq:bigfnu_circle}), in the physical domain, and the poles are
inside the circle, namely, in the unphysical domain (compare with
the Hermitian case, Fig. \ref{fig:bigfnu_dispersion_args}). The black
arrows represent the contour used in (c)--(d): Straight lines along
$k_{y}\approx0.01$ outside the ellipse, glued to an $\approx89^{\circ}$
arc along the upper half-circle with radius $\approx1.064\cdot K_{c}$,
so it passes \textquotedblleft above\textquotedblright{} the roots,
namely, with $k_{y}>K_{y}$. (b) Bulk and edge dispersions. Colors
are as in Fig. \ref{fig:interpolated_phase}. The horizontal blue
lines and light-blue shaded areas indicate bulk dispersion projection
onto the plane $\left(k_{x},\omega\right)$ from the three-dimensional
Mexican hat shape, filling the convex hull. We see three edge mode
dispersions emerging from the bulk dispersion. One is the parameter-independent
$\omega=-k_{x}$ emerging outside the ellipse, and the two other modes
emerge inside the ellipse, with $K_{y}\protect\ne0$, and away from
the bottom of the band along the circle. (c) The phase accumulation
of $\det S$ along the contour depicted in (a). Dashed lines represent
the phase of the scalar $-S_{\infty}$ outside the ellipse, and continuous
lines represent the phase of $\det S$ inside. The phase increases
by $2\pi$ for each mode emergence in a localized manner along this
contour. Additional phase changes of $\pm\pi$ are seen at the edges
of the ellipse. These artifacts of the gluing between regimes cancel
out and do not affect the BEC. (d) The magnitude of $\det S$ along
the same contour. It approaches zero around the edge-bulk dispersion
intersections inside the ellipse, for which the root and pole are
separated. The vertical lines in (b)--(d) indicate the $k_{x}$ values
at which the edge modes emerge (see above). (f) The phase accumulation
of the scattering matrix for $|\mathbf{k}|\to\infty$, using $S_{0}$,
along a straight line contour with large constant $k_{y}$. As before
(see Fig. \ref{fig:adep_phase}), since the contour direction is inverted,
the phase increase of $+2\pi$ must be interpreted as $-2\pi$, which,
together with the $3\times2\pi$ contribution of the $3$ edge modes
in (d), leads to the generalized BEC, $\left(6\pi-2\pi\right)/2\pi=2=C_{+}$.}
\end{figure*}

\subsection{Non-Hermitian edge problem\label{subsec:bigfnu-Non-Hermitian}}

In this final case we consider both large odd viscosity, which leads
to a $2\times2$ scattering matrix (\ref{eq:bigfnu_Smatrix}) inside
the ellipse (\ref{eq:bigfnu_ellipse}), and the non-Hermitian partial-slip
boundary condition (\ref{eq:explicit_interpolated_BC}), which leads
to edge modes decaying in time, $\im{\omega}<0$. The results show
the combined effects of these two modifications. We observe the separation
of the roots and poles of the scattering matrix ($\det S$) away from
the minimum of the band, that is, the circle (\ref{eq:bigfnu_circle}).
The dependence on the parameter $\tilde{a}$ remains as in Sec. \ref{sec:Non-hermitian-edge-problem}
--- for $\tilde{a}=0$ there are two finite-$k_{x}$ edge modes emerging
from the bottom of the bulk band, and for $\tilde{a}>0$ there are
three finite-$k_{x}$ edge modes, where the featureless $\omega=-k_{x}$
mode remains as before, whereas the other two modes now emerge outside
the circle, that is, away from the bottom of the band. Moreover, the
topological structure remains as before, namely the generalized BEC
is obeyed with the use of the non-Hermitian relative Levinson's theorem
(\ref{eq:modified_levinson_thm_orr}), i.e., we take a contour ``above''
the roots in order for $\arg\left(\det S\right)$ to count the number
of edge-bulk dispersion intersections correctly.

Similarly to the previous case, the calculation of the scattering
matrix inside the ellipse (\ref{eq:bigfnu_ellipse}) results in the
form given in Eq. (\ref{eq:bigfnu_Smatrix}), with the difference
only in $G$, which is now given by
\begin{widetext}
\begin{equation}
\tilde{G}\left(k_{x},k_{y,1},k_{y,2}\right)=\begin{vmatrix}\left(i\tilde{a}\nu k_{x}-1\right)\hat{u}_{\infty}\left(k_{x},k_{y,1}\right)-i\tilde{a}\nu k_{y,1}\hat{v}_{\infty}\left(k_{x},k_{y,1}\right) & \left(i\tilde{a}\nu k_{x}-1\right)\hat{u}_{\infty}\left(k_{x},k_{y,2}\right)-i\tilde{a}\nu k_{y,2}\hat{v}_{\infty}\left(k_{x},k_{y,2}\right)\\
\hat{v}_{\infty}\left(k_{x},k_{y,1}\right) & \hat{v}_{\infty}\left(k_{x},k_{y,2}\right)
\end{vmatrix}.\label{eq:GforS_interpolated_bigfnu}
\end{equation}
\end{widetext}

\noindent Direct calculation shows the determinant is of the same
form as before {[}see Eq. (\ref{eq:bigfnu_hermitian_detS}){]}, but
with $\tilde{G}$ instead of $G$. Here $\det S$ is not pure phase,
and does not have the second property in Eq. (\ref{eq:bigfnu_hermitian_Gproperties}).
In Appendix \ref{sec:app_detS=00003D1_bigfnu_interpolated} we show
analytically that $\left|\det S\right|<1$ in the physical domain
(inside the ellipse and outside the circle with positive $k_{y,1}$),
as expected from dissipative (as opposed to gain) boundary conditions.
This is supported by our numerical calculations, which further show
that $S$ has two distinct eigenvalues $\lambda_{j},j=1,2$. These
obey $\left|\lambda_{j}\right|\leq1$ in the physical domain, with
equality achieved along the circle (\ref{eq:bigfnu_circle}). Hence
$S$ is always diagonalizable in our system, even though it is no
longer unitary. 

Similarly to the case of small odd viscosity in Sec. \ref{subsec:interpolated-generalized-BEC},
we can again see, using the continuity relations (\ref{eq:bigfnu_continuity_kykappa}),
that the equation for edge dispersion at these points is exactly the
same as requiring $\det S=0$. Indeed, we see (Fig. \ref{fig:interpolated_bigfnu_dispersion_args})
that again roots and poles are separated, such that the roots are
in the physical domain (outside the circle) and the poles are in the
unphysical domain (inside the circle). This is also illustrated in
Appendix \ref{sec:app-Topographic-map-of_dets}.

As in the previous case, we glue two contours: (i) Outside and away
from the ellipse a straight line contour along the bottom of the band.
(ii) Inside the ellipse and close to it outside, for an edge mode
dispersion that intersects the bulk band at $\left(K_{x},K_{y}\right)$
we take $R^{2}=\left(K_{x}\right)^{2}+\left(K_{y}\right)^{2}+\epsilon$
and evaluate $\arg\left(\det S\right)$ along $k_{x}^{2}+k_{y}^{2}=R^{2}$
with $k_{y}>K_{y}>0$. Our numerical results indeed show the phase
of $\det S$ accumulated along this path inside the ellipse counts
the edge modes according to the relative Levinson's theorem, and the
winding along the top of the bulk band recovers the generalized BEC,
see Fig. \ref{fig:interpolated_bigfnu_dispersion_args}. 

\section{Discussion and conclusions \label{sec:Discussion-and-conclusions}}

In this work we have expanded our understanding of the generalized
BEC to the simplest non-Hermitian case, where the bulk problem is
Hermitian, and non-Hermiticity arises only from the boundary condition
we have introduced. We have seen that the relative Levinson's theorem
can be generalized so as to apply even in this case, while the behavior
at infinite wave vector was not affected, allowing us to adjust the
generalized BEC to systems with non-Hermitian boundary conditions.
We have also allowed large odd viscosity, where the scattering matrix
becomes a $2\times2$ matrix in some parts of the wavevector plane.
Upon suitably generalizing the definition of phase of the scattering
matrix, we found it also obeys the same form of generalized BEC, even
with the non-Hermitian boundary condition.

In future works we intend to use the partial-slip boundary condition
together with a non-Hermitian bulk problem, by adding even (dissipative)
viscosity as well as friction terms to Eqs. (\ref{eq:orig_PDEs}).
There one can expect exceptional rings to affect the BEC \citep{Zhen2015,Bergholtz2021,Souslov2019}.
Further down the road we envision taking this model back to the context
of solid-state systems, and specifically for edge magnetoplasmons
emerging in two-dimensional electron gas under the influence of a
magnetic field, where the long-range Coulomb interaction presents
the main challenge \citep{Cohen2018b}. Another possible route for
the future, in order to model more realistic systems, is to take into
account the effects of disorder \citep{Burmistrov2019}, on top of
non-Hermiticity. The tools developed here may also be relevant to
testing the BEC in dissipation-induced topological states \citep{Goldstein2019,Shavit2020,Beck2021}.
Systems in other topological classes (beyond Chern insulators) \citep{Hasan2010,Bansil2016,Qi2011}
would also be interesting to study.
\begin{acknowledgments}
Our work has been supported by the Israel Science Foundation (ISF)
and the Directorate for Defense Research and Development (DDR\&D)
grant No. 3427/21 and by the US-Israel Binational Science Foundation
(BSF) Grant No. 2020072.
\end{acknowledgments}

\appendix

\section{Negative odd viscosity obeys the generalized BEC \label{sec:App_neg_oddvisc}}

The case of $f\nu<0$ obeys the same generalized BEC in both Hermitian
and non-Hermitian cases. Assume $f>0,\,\nu<0$ (the case $f<0,\,\nu>0$
has the same topological structure, but with inverted signs). The
bulk eigenmodes in Eq. (\ref{eq:psi_inf_sol_q}) are regular on $\ar^{2}\cup\left\{ \infty\right\} $,
since $q\rightarrow1$ for $|\mathbf{k}|\to\infty$, independently
of the angle, which implies trivial bulk topology, in agreement with
the bulk topological invariant being $C_{+}=0$, which was calculated
in Refs. \citep{Tauber2018,Souslov2019} {[}see Eq. (\ref{eq:Chern_in_SWM}){]}.
Note that there is no need to move a singularity of the bulk eigenmodes,
so $S_{\infty}$ is used for the calculations along the top of the
upper band as well.

Similarly to the main text, our numerical calculations give the dependence
of the number of finite-$\mathbf{k}$ edge modes emerging from $(+)$
or merging into $(-)$ the upper bulk band on the parameters of the
boundary conditions: (i) For the Hermitian boundary condition (\ref{eq:boundary_cond})
we find $0,-1,1,0$ edge modes for $a\in\left(-\infty,-\sqrt{2}\right),\left(-\sqrt{2},0\right),\left(0,\sqrt{2}\right),\left(\sqrt{2},\infty\right)$,
respectively. (ii) For the non-Hermitian boundary condition (\ref{eq:interpolated_BC})
we find $0,-1,-1$ edge modes for $\tilde{a}=0,\tilde{a}\in\left(0,\infty\right),\tilde{a}=\infty$,
respectively. The apparent mismatch with the bulk Chern number $C_{+}$
is amended by the behavior of the scattering matrix at $|\mathbf{k}|\to\infty$,
see Eq. (\ref{eq:NH_generalized_BEC}). Note that the sign of the
term $\mathrm{Ind}\left(S\left(\left|\mathbf{k}\right|\rightarrow\infty\right)\right)$
is inverted here with relation to the case of positive $f,\nu$, and
that the parameter independent \textcolor{red}{Kelvin wave} solution,
$\omega=-k_{x}$, $\emph{does not }$occur.

\section{Proof of the Hermiticity of the boundary condition (\ref{eq:boundary_cond})
and the dissipative nature of the boundary condition (\ref{eq:interpolated_BC})\label{sec:App.proof-NHBCisdissipative}}

In this appendix we prove that the Hermitian boundary conditions (\ref{eq:boundary_cond})
indeed preserve the Hermiticity of the edge problem for any $a\in\ar$
(as was already shown in Ref. \citep{Graf2020}, in a less physically-motivated
approach), whereas the non-Hermitian partial-slip boundary conditions
(\ref{eq:interpolated_BC}) break Hermiticity for any finite $\tilde{a}$,
and specifically lead to loss and not gain. 

Consider $H$, the Hamiltonian in the main text, Eq. (\ref{eq:orig_hamiltonian})
and vectors in the edge problem, $\phi,\psi\in L_{2}\left(\ar\times\ar_{+}\right)^{\otimes3}$,
i.e., on the half-plane $\left(x,y\right)\in\ar\times\ar_{+}$. Both
boundary conditions we consider include the condition 
\begin{equation}
\varphi_{3}|_{y=0}=0,\label{eq:common_condition_BCs}
\end{equation}
 and from the definition of the $L_{2}$ space we have the additional
assumptions 
\begin{equation}
\lim_{x\rightarrow\pm\infty}\varphi_{i}=\lim_{y\rightarrow+\infty}\varphi_{i}=0,\label{eq:L2_vanishatinf}
\end{equation}
for $i=1,2,3$ and for all $\varphi\in L_{2}\left(\ar\times\ar_{+}\right)^{\otimes3}$.
We first prove the Hermiticity of the edge problem, namely, $\langle\phi|H|\psi\rangle=\langle\psi|H|\phi\rangle^{*}$,
for the boundary condition (\ref{eq:boundary_cond}), for all $a\in\ar$.
Direct calculation, using integration by parts and the assumptions
(\ref{eq:common_condition_BCs}) and (\ref{eq:L2_vanishatinf}), leads
to
\begin{widetext}
\begin{multline}
\langle\phi|H|\psi\rangle=\int_{0}^{\infty}dy\int_{-\infty}^{\infty}dx\left\{ \psi_{2}^{*}(-i\partial_{x})\phi_{1}+\psi_{1}^{*}(-i\partial_{x})\phi_{2}+\psi_{3}^{*}(-i\partial_{y})\phi_{1}+\psi_{1}^{*}(-i\partial_{y})\phi_{3}\right.\\
\left.+\psi_{3}^{*}\left[i\left(f+\nu\nabla^{2}\right)\right]\phi_{2}+\psi_{2}^{*}\left[-i\left(f+\nu\nabla^{2}\right)\right]\phi_{3}\right\} ^{*}-i\nu\int_{-\infty}^{\infty}dx\left[\phi_{2}^{*}\partial_{y}\psi_{3}+\psi_{2}\partial_{y}\phi_{3}^{*}\right]_{y=0}^{\infty}\\
=\langle\psi|H|\phi\rangle^{*}+i\nu\int_{-\infty}^{\infty}dx\left(\phi_{2}^{*}\partial_{y}\psi_{3}+\psi_{2}\partial_{y}\phi_{3}^{*}\right)|_{y=0},\label{eq:HermitianHexplicit}
\end{multline}
\end{widetext}

\noindent where specifically the boundary terms resulting from the
density gradient (e.g. $\phi_{2}^{*}\do_{x}\psi_{1}$) and Coriolis
(e.g. $\phi_{2}^{*}f\psi_{3}$) vanish. Since the boundary condition
(\ref{eq:boundary_cond}) implies $\partial_{y}\varphi_{3}=-a^{-1}\partial_{x}\varphi_{2}$,
the remaining boundary term, related to the odd viscosity, vanishes
as well
\begin{multline}
i\nu\int_{-\infty}^{\infty}dx\left(\phi_{2}^{*}\partial_{y}\psi_{3}+\psi_{2}\partial_{y}\phi_{3}^{*}\right)|_{y=0}\\
=-i\frac{\nu}{a}\int_{-\infty}^{\infty}dx\left(\phi_{2}^{*}\partial_{x}\psi_{2}+\psi_{2}\partial_{x}\phi_{2}^{*}\right)|_{y=0}\\
=-i\frac{\nu}{a}\int_{-\infty}^{\infty}dx\partial_{x}\left(\phi_{2}^{*}\psi_{2}\right)|_{y=0}=0.
\end{multline}
So indeed $\langle\phi|H|\psi\rangle=\langle\psi|H|\phi\rangle^{*}$
for all $a\in\ar$. Recall that $a=-1$ is the no-stress condition,
so we have proved its Hermiticity along the way as well. Furthermore,
we note that the above calculation also shows that the no-slip condition
($\varphi_{2}|_{y=0}=0$) leads to a Hermitian edge problem, since
the last term in Eq. (\ref{eq:HermitianHexplicit}) vanishes trivially.

For the partial-slip boundary conditions (\ref{eq:interpolated_BC})
with $\tilde{a}>0$, we consider the expectation value, $\langle\psi|H_{1}|\psi\rangle$,
of the operator $H_{1}=\left(H-H^{\dagger}\right)/2i=\im H$, and
show that it is strictly non-positive, which implies that the operator
$H_{1}$ is negative-semidefinite. In fact, we will show that the
edge frequency eigenvalues of $H$ have negative imaginary part, $\im{\omega}<0$,
which leads to edge modes decaying in time (except for the boundary-condition
independent $\omega=-k_{x}$ edge mode, for which $v=0$ everywhere
so the second term in Eq. (\ref{eq:HermitianHexplicit}) vanishes,
which means it is Hermitian). Thus, the edge problem with the partial-slip
boundary conditions has strictly non-real spectrum, so it is non-Hermitian.
Again using the assumptions (\ref{eq:common_condition_BCs}),(\ref{eq:L2_vanishatinf}),
we see that the density gradient and Coriolis terms do not contribute
to the expectation value $\langle\psi|H_{1}|\psi\rangle$, since all
the corresponding boundary terms vanish upon using integration by
parts. Hence, we may focus solely on the terms involving odd viscosity.
The viscous terms appear in the Hamiltonian as $\left(H\mathbf{v}\right)_{a}=-i\left(\nabla\cdot T\right)_{a}$,
with $\mathbf{v}=\left(u,v\right)^{T}$, so we may write (employing
Einstein summation)
\begin{multline}
\langle\psi|H_{1}|\psi\rangle=\im{\langle\psi|H|\psi\rangle}\\
=\im{-i\int_{0}^{\infty}dy\int_{-\infty}^{\infty}dxv_{a}^{*}\partial_{b}T_{ab}}\\
=-\re{\int_{0}^{\infty}dy\int_{-\infty}^{\infty}dxv_{a}^{*}\partial_{b}T_{ab}}.
\end{multline}
Using integration by parts and the assumptions above, we find
\begin{multline}
\langle\psi|H_{1}|\psi\rangle=\mathrm{Re}\left[\int_{0}^{\infty}dy\int_{-\infty}^{\infty}dx\left(\partial_{b}v_{a}^{*}\right)T_{ab}\right]\\
+\mathrm{Re}\left[\int_{-\infty}^{\infty}\left(u^{*}T_{xy}\right)|_{y=0}dx\right].\label{eq:H1_twoterms}
\end{multline}
The first term in the above expression vanishes, since we only consider
odd viscosity: Substituting $T_{ab}=-\nu_{abcd}^{-}\partial_{d}v_{c}$,
we observe that 
\begin{multline}
\mathrm{Re}\left[\int_{0}^{\infty}dy\int_{-\infty}^{\infty}dx\left(\partial_{b}v_{a}^{*}\right)T_{ab}\right]\\
=-\mathrm{Re}\left[\int_{0}^{\infty}dy\int_{-\infty}^{\infty}dx\nu_{abcd}^{-}\left(\partial_{b}v_{a}^{*}\right)\left(\partial_{d}v_{c}\right)\right],
\end{multline}
and note that the term $\nu_{abcd}^{-}\left(\partial_{b}v_{a}^{*}\right)\left(\partial_{d}v_{c}\right)$
is pure imaginary, since, by definition, the odd part satisfies $\nu_{abcd}^{-}=-\nu_{cdab}^{-}$
\citep{Avron1998}. For any $\tilde{a}>0$, we plug in the boundary
condition (\ref{eq:interpolated_BC}), $u|_{y=0}=-\tilde{a}T_{xy}|_{y=0}$,
and find that the second term in Eq. (\ref{eq:H1_twoterms}) is indeed
negative
\begin{equation}
\langle\psi|H_{1}|\psi\rangle=-\frac{1}{\tilde{a}}\mathrm{Re}\left[\int_{-\infty}^{\infty}\left(u^{*}u\right)|_{y=0}dx\right]<0.
\end{equation}
Thus, as expected in the main text, $\tilde{a}>0$ leads to dissipation
of all edge modes {[}except for the $\omega=-k_{x}$ one, for which
$u=0$ at the edge for all $\tilde{a}$, see Eq. (\ref{eq:explicit_interpolated_BC}){]},
while $\tilde{a}<0$ would be unphysical, as it leads to gain ($\im{\omega}>0$)
in the edge problem. We briefly note that the Hermitian boundary conditions
obey $\langle\psi|H_{1}|\psi\rangle=0$, but of course this is not
enough to prove Hermiticity, since non-Hermitian operators may have
strictly real spectrum as well, hence above we resorted to a direct
proof of Hermiticity.

\section{The scattering matrix decreases amplitudes in the non-Hermitian case\label{sec:App-detS=00003D1_interpolated_bc}}

In this appendix we show that for a $1\times1$ scattering matrix,
$\left|S_{\zeta}\left(k_{x},k_{y}\right)\right|<1$ for $k_{y}>0$,
which corresponds to the boundary condition (\ref{eq:explicit_interpolated_BC})
causing dissipation (i.e., loss rather than gain), in the domain where
it is well-defined in terms of direction of wave propagation. Note
that here, as in the Hermitian case, moving the singularity does not
affect the absolute value of the scattering matrix, since 
\begin{equation}
S_{\zeta}\left(k_{x},k_{y}\right)=\frac{t_{\infty}^{\zeta}\left(k_{x},-k_{y}\right)}{t_{\infty}^{\zeta}\left(k_{x},k_{y}\right)}S_{\infty}\left(k_{x},k_{y}\right).
\end{equation}
Since, by Eq. (\ref{eq:tzeta}), $\left|t_{\infty}^{\zeta}\left(k_{x},k_{y}\right)\right|=1$,
it is indeed enough to consider $S_{\infty}$. By continuity, it is
enough to show that $\left|S_{\infty}\right|=1$ if and only if $k_{y}=0$,
using the fact that we have seen the roots (where $\left|S_{\infty}\right|=0<1$)
are in $k_{y}>0$. One direction is trivial: plug $k_{y}=0$ and see
$S_{\infty}=-1$. For the other direction, we write the equation
\begin{equation}
\left|S_{\infty}\right|^{2}=1,
\end{equation}
and assume $k_{y}\ne0$. We expand using Eq. (\ref{eq:interpolated_Smatrix})
and the definitions of the bulk modes (\ref{eq:psi_inf_sol_q}) to
find
\begin{multline}
0=\nu\omega_{+}\im{\kev}\mathbf{k}^{2}\left(2\mathbf{k}^{2}+\frac{1-2f\nu}{\nu^{2}}\right)+\\
+k_{x}\mathbf{k}^{2}\left[2\nu^{2}\mathbf{k}^{4}+\mathbf{k}^{2}\left(3-4f\nu\right)\right.\\
\left.+\left(-\frac{f}{\nu}+\frac{1}{\nu^{2}}\left(1-2f\nu\right)+2f^{2}\right)\right].\label{eq:appA_eq_|S|=00003D1}
\end{multline}
Working under the assumption that $f\nu<1/4$ we see that the first
term is strictly positive. The second term vanishes only at
\begin{equation}
\mathbf{k}_{\pm}^{2}=\frac{1}{\nu^{2}}\left(f\nu+\frac{-3\pm1}{4}\right)<0,
\end{equation}
so it is strictly positive as well. Therefore, for any $\mathbf{k}^{2}\ne0$,
the entire RHS in Eq. (\ref{eq:appA_eq_|S|=00003D1}) is positive,
and indeed there are no other solutions to the equation.

\begin{figure*}[t]
\includegraphics[scale=0.6]{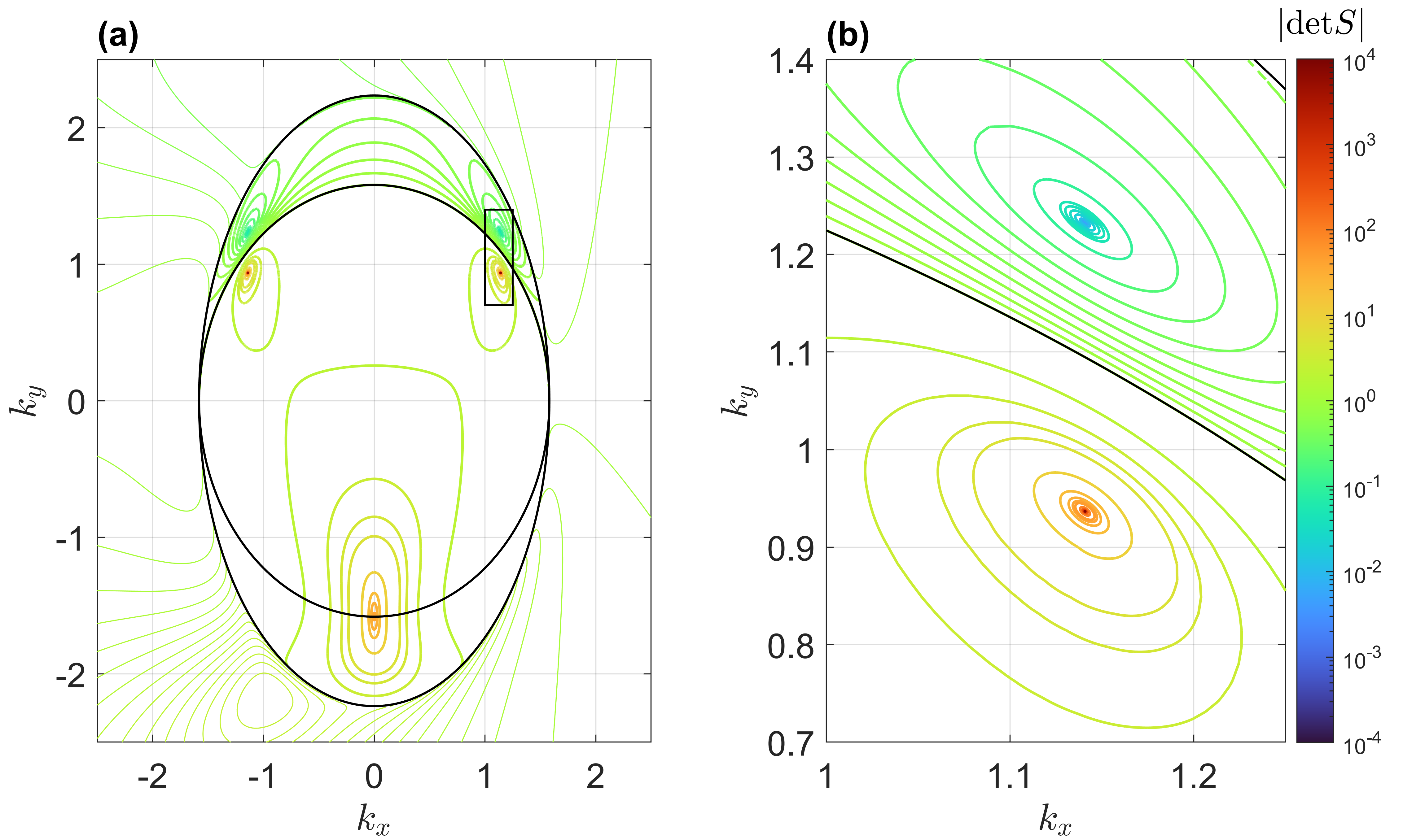}\caption{\label{fig:Topographic-map-of-detS} (a) Contour plot of $\left|\det S\right|$
for large $f\nu$, with the non-Hermitian boundary condition (\ref{eq:explicit_interpolated_BC}).
Here $f=3,\nu=1,\tilde{a}=2$, as in Fig. \ref{fig:interpolated_bigfnu_dispersion_args}.
Inside the ellipse the scattering matrix is $2\times2$ and outside
it is a scalar, with continuity along the ellipse between $\det S$
inside and $S_{\infty}$ outside, up to a sign correction {[}see Eq.
(\ref{eq:bigfnu_Scontinuity_ellipse}){]}. We clearly see the two
roots at $\left(K_{x},K_{y}\right)\approx\left(\pm1.14,1.23\right)$,
outside the circle, and the two poles at $\left(K_{x},K_{y}\right)\approx\left(\pm1.14,0.94\right)$,
inside the circle (up to the plot's finite resolution). In contrast,
the hill on the negative part of the $k_{y}$ axis, in the unphysical
domain, has a finite height and is thus not a pole. As mentioned in
the text, in the physical domain (the upper half-plane outside the
ellipse, and outside the circle with positive $k_{y}$ inside the
ellipse), we have $\left|\det S\right|<1$, as expected from dissipative
boundary conditions. (b) Zoom in of (a) on a root and pole pair. The
same colorbar applies to both panels.}
\end{figure*}

\section{The anomaly of Levinson's theorem at infinity for the non-Hermitian
case\label{sec:APP-Analytic-continuation}}

Similarly to Sec. 3.5 of Ref. \citep{Graf2020}, we expand $S_{0}$
{[}Eq. (\ref{eq:interpolated_Smatrix}){]} in the non-Hermitian case
at $\left|\mathbf{k}\right|\rightarrow\infty$, and show an anomaly
of Levinson's theorem occurs for all $\tilde{a}>0$ but not for the
no-slip condition, $\tilde{a}=0$. Even though the anomaly is absent
from the $\tilde{a}=0$ case, we find the winding of $S_{0}$ at infinity
to be in agreement with our numerical results in the main text.

First, we briefly explain the anomaly as presented in Ref. \citep{Graf2020}.
There it is shown for the family of boundary conditions (\ref{eq:boundary_cond}),
that the winding of the scattering matrix along the top of the upper
bulk band, corresponds to no edge mode at $\left|\mathbf{k}\right|\rightarrow\infty$.
More explicitly, this means that for $\left|a\right|<\sqrt{2}$, where
the BEC is amended by the winding of the scattering matrix along the
top of the band, there are no edge modes at $\left|\mathbf{k}\right|\rightarrow\infty$,
in the sense that there are no poles of $S_{0}$ there. The opposite
is true for $\left|a\right|>\sqrt{2}$, namely, while the edge index
of the finite-$\mathbf{k}$ edge modes matches the Chern number, and
accordingly there is no winding of the scattering matrix at $\left|\mathbf{k}\right|\rightarrow\infty$,
there are edge modes there (poles of $S_{0}$). We will now show that
with the non-Hermitian boundary conditions (\ref{eq:explicit_interpolated_BC}),
similar behavior occurs for $\tilde{a}>0$, where $S_{0}$ winds at
infinity and there are no edge modes there, but not for $\tilde{a}=0$,
where there are neither winding nor edge modes.

For the non-Hermitian boundary condition (\ref{eq:explicit_interpolated_BC}),
we expand $g_{0}\left(k_{x},-k_{y}\right)$ {[}see Eq. (\ref{eq:interpolated_Smatrix}){]}
at $\left|\mathbf{k}\right|\rightarrow\infty$ and find
\begin{equation}
g_{0}\left(k_{x},-k_{y}\right)\sim\frac{i}{2}\left[\tilde{a}\nu\left(2ik_{x}+k_{y}+\kev\right)-2\right],\label{eq:app_g0expansion}
\end{equation}
\begin{equation}
\arg\left(S_{0}\right)=2\arg\left(g_{0}\left(k_{x},-k_{y}\right)\right),\label{eq:app_args0tog0}
\end{equation}
 and from Eq. (\ref{eq:kappa_ev_def}),
\begin{equation}
\kev\sim i\sqrt{2k_{x}^{2}+k_{y}^{2}}.\label{eq:app_kev_atinf}
\end{equation}
For $\tilde{a}>0$ the leading order expansion is thus 
\begin{equation}
g_{0}\left(k_{x},-k_{y}\right)\sim\frac{i}{2}\left[\tilde{a}\nu\left(2ik_{x}+k_{y}+\kev\right)\right].\label{eq:app_g0_at_kinf}
\end{equation}
Next, we take the reciprocal coordinates 
\begin{equation}
k_{x}=\frac{\lambda_{x}}{\lambda_{x}^{2}+\lambda_{y}^{2}},k_{y}=\frac{\lambda_{y}}{\lambda_{x}^{2}+\lambda_{y}^{2}},\label{eq:recip_coordinates}
\end{equation}
which exchange $0$ and $\infty$, and thus, using Eq. (\ref{eq:app_kev_atinf}),
we get 
\begin{equation}
g_{0}\left(k_{x},-k_{y}\right)\sim-\frac{\tilde{a}\nu}{2\left(\lambda_{x}^{2}+\lambda_{y}^{2}\right)}\left(2\lambda_{x}+\sqrt{2\lambda_{x}^{2}+\lambda_{y}^{2}}-i\lambda_{y}\right).\label{eq:app_g0_expansion_recipcoor}
\end{equation}
From this expression we now show that $S_{0}$ does wind at $\left|\mathbf{k}\right|\rightarrow\infty$,
and yet, there are no roots of $S_{0}$ there. For the calculation
of the winding of $S_{0}$, the prefactor in Eq. (\ref{eq:app_g0_expansion_recipcoor})
is irrelevant, while the last term, $-i\lambda_{y}$, determines that
the phase will be accumulated from below the real line. Futhermore,
for any (small) constant $k_{y}>0$ we have 
\begin{equation}
\lim_{\lambda_{x}\rightarrow\pm\infty}2\lambda_{x}+\sqrt{2\lambda_{x}^{2}+\lambda_{y}^{2}}=\pm\infty.
\end{equation}
We note that in the parametrization Eq. (\ref{eq:recip_coordinates}),
$\lambda_{x}$ traverses the real line in the opposite direction to
$k_{x}$. Therefore, as $k_{x}$ varies from $-\infty$ to $\infty$,
the phase $\arg\left(g_{0}\left(k_{x},-k_{y}\right)\right)$ is changed
from from $0$ to $-\pi$. Finally, with (\ref{eq:app_args0tog0})
we obtain $\Delta\left[\arg\left(S_{0}\right)\right]=-2\pi$, i.e.,
$\mathrm{Ind}\left(S_{0}\left(k_{y}\rightarrow\infty\right)\right)=-1$,
as expected from the numerical results presented in the main text,
see Fig. \ref{fig:interpolated_phase}. 

For the roots of $S_{0}$, we again use the expansions (\ref{eq:app_g0_at_kinf})
and (\ref{eq:app_kev_atinf}), along with the expansion of Eq. (\ref{eq:original_dispersion})
\begin{equation}
\omega\sim\nu\left(k_{x}^{2}+k_{y}^{2}\right)\sim\nu\left(k_{x}^{2}+\kev^{2}\right).\label{eq:app_omega_atinf}
\end{equation}
Since $g_{0}$ is homogeneous in $k_{x},k_{y}$ {[}see Eq. (\ref{eq:app_g0_at_kinf}){]},
we may use the ansatz
\begin{equation}
k_{y}=ic_{+}k_{x},\,\kev=-ic_{-}k_{x}.
\end{equation}
Thus, from requiring $g_{0}=0$ we have the equation
\begin{equation}
2+c_{+}-c_{-}=0,
\end{equation}
while from Eq. (\ref{eq:app_kev_atinf}) we have the constraint
\begin{equation}
c_{+}^{2}+c_{-}^{2}=2.
\end{equation}
Their only joint solution is $c_{+}=-1,c_{-}=1$, but this is not
a valid solution, since plugging it into Eq. (\ref{eq:app_omega_atinf})
leads to $\omega=0$, which is not in the upper bulk band. So again,
as expected, there is no root of $S_{0}$ at $\left|\mathbf{k}\right|\rightarrow\infty$.

Finally we consider the expansion of $g_{0}$ for $\tilde{a}=0$.
From (\ref{eq:app_g0expansion}) we see that $g_{0}\sim-i$ to leading
order. Thus, there is no winding of $S_{0}$, and also no roots of
$S_{0}$, in contrast to the previous case. We note that higher orders
in the expansion will have lower powers of $\mathbf{k}$, so they
cannot add roots. 

\section{Equivalence of numerical methods for evaluation of the phase of the
scattering matrix\label{sec:App-Equivalence-of-numerical-methods}}

We briefly introduce the two methods for numerical calculation of
the phase of the scattering matrix, which apply to both matrix and
scalar $S$ for both the Hermitian and non-Hermitian cases, and prove
they are equivalent. The first method is to evaluate directly the
phase of the complex number $\varphi=\arg\left(\det S\right)$. This
method is straightforward, but when the phase winds over more than
one $2\pi$ cycle, one must ``unwrap'' the phase accordingly. The
second option is to numerically integrate over $d\varphi=\mathrm{Im}\left(\mathrm{Tr}\left(S^{-1}dS\right)\right)$,
which by nature results in relative phase, and thus its value depends
on the starting point. Since we are interested in the phase difference
accumulated along some finite contour, this is not a problem. 

Let us show in a general framework why these two methods are equivalent:
For all matrices it is known that $\ln\left(\det S\right)=\mathrm{Tr}\left(\ln S\right)$.
Also, for every complex number $\arg\left(z\right)=\im{\ln\left(z\right)}$
(assuming main branch). We now define the phase of a matrix $S$ to
be $\varphi=\mathrm{Im}\left(\mathrm{Tr}\left(\ln S\right)\right)$,
so $\varphi=\im{\ln\left(\det S\right)}=\arg\left(\det S\right)$.
On the other hand, $d\varphi=\im{\mathrm{Tr}\left(d\left(\ln S\right)\right)}\overset{(*)}{=}\im{\mathrm{Tr}\left(S^{-1}dS\right)}$,
with $(*)$ relying on the cyclic property of the trace. Thus we get
$\arg\left(\det S\right)=\int\im{\mathrm{Tr}\left(S^{-1}dS\right)}$,
as wanted. Note that this is true even if $S$ does not commute with
itself at different points along the integration contour, and even
if $S$ is not unitary (which is especially important in the non-Hermitian
case).

\section{The scattering matrix decreases amplitudes in the non-Hermitian case
with large odd viscosity \label{sec:app_detS=00003D1_bigfnu_interpolated}}

Similar to the calculation in Appendix \ref{sec:App-detS=00003D1_interpolated_bc},
we now show that $\left|\det S\right|<1$ for $k_{y}>0$ inside the
ellipse (\ref{eq:bigfnu_ellipse}) and outside the circle (\ref{eq:bigfnu_circle}),
for the $2\times2$ scattering matrix in the non-Hermitian case. It
is enough from continuity to show that $\left|\det S\right|=1$ if
and only if $k_{y,1}=k_{y,2}$, since the solution to the equation
$k_{y,2}\left(k_{y,1}\right)=k_{y,1}$ is exactly the circle, and
we have shown above that the roots (where $\left|\det S\right|=0<1$)
are outside the circle. As before, one direction is trivial --- plugging
$k_{y,1}=k_{y,2}$ into Eqs. (\ref{eq:bigfnu_hermitian_detS}),(\ref{eq:GforS_interpolated_bigfnu})
yields $\det S=-1$. For the other direction we expand the equation
\begin{equation}
\left|\det S\right|^{2}=1
\end{equation}
 to find
\begin{equation}
0=-\left(1-\frac{k_{x}^{2}}{\omega^{2}}\right)\tilde{a}\nu\left(k_{y,1}+k_{y,2}\right)\left(q\left(k_{y,2}\right)-q\left(k_{y,1}\right)\right).
\end{equation}
Using Eq. (\ref{eq:psi_inf_sol_q}) we see
\begin{equation}
q\left(k_{y,2}\right)-q\left(k_{y,1}\right)=\frac{\nu}{\omega}\left(k_{y,1}^{2}-k_{y,2}^{2}\right),
\end{equation}
so indeed the equation
\begin{equation}
0=-\left(1-\frac{k_{x}^{2}}{\omega^{2}}\right)\tilde{a}\nu\left(k_{y,1}+k_{y,2}\right)^{2}\frac{\nu}{\omega}\left(k_{y,1}-k_{y,2}\right)
\end{equation}
has a single solution $k_{y,1}=k_{y,2}$ (and $\omega=-k_{x}$ which
is irrelevant for a scattering state solution).

\section{$\left|\det S\right|$ in the non-Hermitian case with large odd viscosity
\label{sec:app-Topographic-map-of_dets}}

In Fig. \ref{fig:Topographic-map-of-detS}, by plotting $\left|\det S\right|$,
we show that the roots of $\det S$ are in the physical domain, namely
outside the circle (\ref{eq:bigfnu_circle}) and inside the ellipse
(\ref{eq:bigfnu_ellipse}), in the upper half-plane, and the poles
are in the unphysical domain, inside the upper half of the circle.
As mentioned in the main text, in the Hermitian case the root and
pole coalesce at the edge of the circle, which is the bottom of the
bulk band, while in the non-Hermitian case they are separated along
the edge of the circle, as the imaginary part of the edge dispersion
allows them to emerge away from the bottom of the bulk band.

Let us note that the hill on the negative part of the $k_{y}$ axis,
seen in Fig. \ref{fig:Topographic-map-of-detS}, has a finite height,
and is thus not a pole for the shown parameters. For other values
of $\tilde{a}$ there are two poles in the unphysical domain, one
inside the circle and one outside, both with negative $k_{y}$ along
the $k_{y}$ axis. They do not correspond to roots or poles in the
physical domain, and thus bear no physical meaning.\nocite{Tauber2022b,Eidelman2004}

\bibliography{NHBEC_rapoport_v3}

\end{document}